\documentclass[10pt,journal,compsoc]{IEEEtran}

\pdfoutput=1

\usepackage[colorlinks,
linkcolor=red,
anchorcolor=blue,
citecolor=green
]{hyperref}

\usepackage{booktabs}
\usepackage{threeparttable} 
\usepackage{multirow}

\usepackage{graphicx} 
\usepackage{float} 
\usepackage{subfigure} 

\usepackage{amsmath}
\usepackage{amsfonts,amssymb}

\usepackage{cuted}

\usepackage{color}
\usepackage{setspace}
\usepackage{picins}

\ifCLASSOPTIONcompsoc
  \usepackage[nocompress]{cite}
\else
  \usepackage{cite}
\fi

\begin{document}

\title{GMSS: Graph-Based Multi-Task Self-Supervised Learning for EEG Emotion Recognition}

\author{ 
	Yang Li,~\IEEEmembership{Member,~IEEE,}
	Ji Chen,
	Fu Li$^*$,~\IEEEmembership{Member,~IEEE,}
	Boxun Fu,  
	Hao Wu,
	Youshuo Ji,
	Yijin Zhou, 
	Yi Niu,
	Guangming Shi,~\IEEEmembership{Fellow,~IEEE,} 
	Wenming Zheng,~\IEEEmembership{Senior Member,~IEEE}

\IEEEcompsocitemizethanks{
	\IEEEcompsocthanksitem Yang Li, Ji Chen, Fu Li, Boxun Fu, Hao Wu, Youshuo Ji, Yijin Zhou, Yi Niu, Guangming Shi are with the Key Laboratory of Intelligent Perception and Image Understanding of Ministry of Education, the School of Artificial Intelligence, Xidian University, Xi’an, 710071, China. (E-mail: fuli@mail.xidian.edu.cn).
	\IEEEcompsocthanksitem Wenming Zheng is with the Key Laboratory of Child Development and Learning Science (Ministry of Education), School of Biological Sciences and Medical Engineering,
	Southeast University, Nanjing, Jiangsu, 210096, China. 
	
	*Corresponding author
	
}


}


\IEEEtitleabstractindextext{%
\begin{abstract}
Previous electroencephalogram (EEG) emotion recognition relies on single-task learning, which may lead to overfitting and learned emotion features lacking generalization. In this paper, a graph-based multi-task self-supervised learning model (GMSS) for EEG emotion recognition is proposed. GMSS has the ability to learn more general representations by integrating multiple self-supervised tasks, including spatial and frequency jigsaw puzzle tasks, and contrastive learning tasks.
By learning from multiple tasks simultaneously, GMSS can find a representation that captures all of the tasks thereby decreasing the chance of overfitting on the original task, i.e., emotion recognition task.
In particular, the spatial jigsaw puzzle task aims to capture the intrinsic spatial relationships of different brain regions.
Considering the importance of frequency information in EEG emotional signals, the goal of the frequency jigsaw puzzle task is to explore the crucial frequency bands for EEG emotion recognition. 
To further regularize the learned features and encourage the network to learn inherent representations, contrastive learning task is adopted in this work by mapping the transformed data into a common feature space.
The performance of the proposed GMSS is compared with several popular unsupervised and supervised methods.
Experiments on SEED, SEED-IV, and MPED datasets show that the proposed model has remarkable advantages in learning more discriminative and general features for EEG emotional signals.
\end{abstract}

\begin{IEEEkeywords}
EEG emotion recognition, multi-task learning, self-supervised learning, graph neural network.
\end{IEEEkeywords}}

\maketitle

\IEEEdisplaynontitleabstractindextext

\IEEEpeerreviewmaketitle

\IEEEraisesectionheading{\section{Introduction}\label{sec:introduction}}

\IEEEPARstart{E}{motion} is close to everyone and plays an important role in our daily lives \cite{b56}. It is a complex and comprehensive psychological and physiological state that can be characterized by behavioral and physiological signals \cite{b57}. Neuroscience research indicates that physiological signals are closer to the source of emotion than behavioral signals\cite{b2}. As a physiological signal, EEG has the advantage of being difficult to disguise and hide compared with behavioral signals,
such as facial expressions and voice\cite{b3}. Moreover, EEG signals significantly benefited from the technological developments in non-invasive EEG recording methods, and are widely used in the research on emotion recognition\cite{b4}\cite{b5}. 
In recent years, emotion recognition has become a research hotspot in human-computer interaction and affective computing \cite{b1}.

A wide variety of methods has been proposed to effectively analyze EEG emotional signals over the past decades. Traditional machine learning methods typically adopt a two-stage model to implement emotional recognition.
For example, Lin et al.\cite{b6} extracted power spectrum density, differential asymmetry power, and rational asymmetry power as features of EEG signals, and then classified them using a support vector machine to study the relationship between emotion and EEG signals.
Jenke et al.\cite{b7} studied and compared the effects of EEG emotion features extracted from the time domain, the frequency domain, and the time-frequency domain on EEG emotion signal recognition.
However, traditional machine learning methods rely on handcrafted features and expert experience \cite{b65}.
With the spectacular success of deep learning methods in the field of computer vision and language recognition, many researchers have considered deep learning models for EEG emotion signals for their ability to automatically extract complex features\cite{b66}\cite{b67}.
For instance, some researchers utilized convolutional neural networks (CNNs) and recurrent neural networks (RNNs) to handle emotion recognition\cite{b8}\cite{b9}. 
Recently, the topological structure of EEG signals has been increasingly studied in EEG emotion recognition owing to the superior performance of graph neural networks (GNNs) in irregular data structures\cite{b10}. 
Wang et al. \cite{b64} proposed a multichannel EEG emotion recognition method based on phase locking value (PLV) graph convolutional neural networks (P-GCNN) to extract the spatio-temporal characteristics and the inherent information in functional connections.

Based on the literature, most EEG-based emotion recognition methods basically face three challenges:
(1) how to generalize the emotion recognition model, and correctly classify new data; (2) how to make full use of EEG characteristics to capture more discriminative data representation for emotion recognition; and (3) how to solve the problem of emotional noise labels.
Regarding the first challenge, EEG displays a highly heterogeneous and nonstationary behavior because emotional signals usually consist of many neural process\cite{b16}. The enormous data distribution shift leads to a lack of generalization for data from different subject or new situation of the current subject\cite{b46}.
Thus, some researchers have adopted the domain adaptation (DA) method to improve generalization. For example, Li et al.\cite{b46} proposed a bi-hemisphere domain adversarial neural network (BiDANN) that contains three domain discriminators to assist with the learning of discriminative emotional features, narrowing the distribution gap between training and testing data, and improving the generality of the recognition model. However, most DA-based methods achieve generality by training the model on labeled training data and unlabeled testing data, which is not suitable for real applications. Thus, it is meaningful and applicable to explore other methods that learn general data representations without test data. 
For the second challenge, 
handcrafted features such as power spectrum density, statistical measure, and discrete wavelet transform are frequently used for generic EEG signal classification tasks. However, these features are not specially designed for EEG emotion signal\cite{b46}.
This issue has also been discussed in recent deep-learning literature on EEG emotion recognition. For example, Zheng et al.\cite{b40} employed a deep belief network (DBN) to directly model the EEG emotion signal. Even though these handcrafted and deep features have been able to extract certain emotion discriminative information, they do not sufficiently exploit specific emotion-related information in EEG emotion recognition tasks. Thus, it is necessary to utilize the characteristics of EEG signals to extract high-level features.
Regarding the third challenge, the emotion labels in the collected EEG data may be noisy and inconsistent as participants may not always produce the expected emotions when watching emotions stimulate stimuli\cite{b58}. Consequently, it is challenging and meaningful to explore how to solve the problem of emotional noise labels that are often ignored in EEG emotion recognition research.

To address the above three major issues in EEG emotion recognition tasks, 
in this paper we propose GMSS, which can learn general EEG emotion representation and improve EEG emotion recognition ability by solving three self-supervised pretext tasks.
To improve generality, GMSS adopts multiple EEG emotion-related tasks that share learned knowledge to generate more general features and avoid overfitting\cite{b49}.
GMSS consists of two graph-based jigsaw puzzle tasks and a contrastive learning task, making it capable to study the impact of emotional expression on spatial and frequency information.
The spatial jigsaw puzzle task enables the predefined distant electrodes to become neighbor electrodes and more emotion-related spatial information is learned in return. 
Meanwhile, the frequency jigsaw puzzle task explores crucial frequency bands for EEG emotion recognition.
Utilizing the augmented samples of the above jigsaw puzzle tasks, the contrastive learning task further standardizes the feature space and enhances the generalization ability of the model. 
These self-supervised pretext tasks, which are based on the intrinsic attributes of EEG emotion data, allow GMSS to deal with EEG noise labels without semantic labeling.
In this study, both unsupervised and supervised approaches of GMSS were evaluated. The experimental results show that GMSS achieves state-of-the-art (SOTA) performance on three public datasets. 

In summary, the contributions of this work can be outlined as follows:
\begin{itemize}
	\item To the best of our knowledge, this is the first work that adopts multi-task learning to improve model generalization capability and avoid overfitting in EEG emotion recognition.
	\item Through the pretext tasks of jigsaw puzzles and contrastive learning, GMSS learns more discriminative features and  alleviates the problem of emotional noise labels, which further improves EEG emotion recognition.
	\item The experimental results, based on both unsupervised and supervised learning approaches, demonstrate that GMSS can achieve SOTA performance on three benchmark datasets.
\end{itemize}

The rest of this paper is organized as follows: Section \uppercase\expandafter{\romannumeral2} provides an overview of previous studies on EEG emotion recognition, graph neural networks, multi-task learning, and self-supervised learning. Section \uppercase\expandafter{\romannumeral3} specifies the GMSS method and its application to EEG emotion recognition. In section \uppercase\expandafter{\romannumeral4} the proposed method is evaluated for EEG emotion recognition through extensive experiments. Finally, section \uppercase\expandafter{\romannumeral5} concludes the paper.

\section{Related Works}
In this section, related works on EEG-based emotion recognition, graph neural networks, multi-task learning, and self-supervised learning are introduced.

\subsection{EEG-based Emotion Recognition}
The general process of EEG emotion recognition includes feature extraction and classification.
Traditional machine learning-based methods typically adopt the statistical measure, discrete wavelet transform, or power spectrum density\cite{b6} as features and then classify the extracted features using SVM, LDA, or LR\cite{b60}. However, deep learning based methods generally extract features by designing feature extraction neural networks followed by linear layers to achieve classification. Many deep learning methods such as CNN, RNN and GNN have been introduced to effectively distinguish different emotional states in EEG emotional signals.
Li et al.\cite{b15} proposed a hierarchical spatial-temporal neural network (R2G-STNN) based on a bidirectional long short-term memory (BiLSTM) network to capture the intrinsic spatial relationships of EEG electrodes within the brain region and between brain regions for EEG emotion recognition.
Song et al.\cite{b11} proposed a multichannel EEG emotion recognition method based on a novel dynamic graph convolutional neural network (DGCNN) to dynamically learn the intrinsic relationship between different EEG channels to assist with features classification.
Zhong et al. \cite{b58} proposed a regularized graph neural network (RGNN) with two regularizers to deal with cross-subject EEG variations and the noise label problem, and achieved promising results.
Li et al.\cite{b13} proposed a bi-hemispheric discrepancy model (BiHDM) to learn discrepancy information between two hemispheres to improve EEG emotion recognition ability. 

\subsection{Graph Neural Network}
The traditional convolutional neural network is excellent for dealing with Euclidean data. 
However, GNN is suitable for handling non-Euclidean data and has shown great promise in the field of social networks, recommendation systems, and knowledge maps\cite{b63}\cite{b61}\cite{b62}.
The GNN fall into two categories, spectral-based and spatial-based. The spectral-based method specifies graphic convolution by introducing a filter from the perspective of graphic signal processing, where the graphic convolution operation is viewed as noise removal from the graphic signal. The spatial-based method is based on the recurrent neural network theory and defines the graph convolution through information propagation\cite{b22}.
Defferrard et al.\cite{b23} argued that the original spectrum convolution suffers from the disadvantages of a large number of parameters and high complexity, and proposed a fast localized convolution algorithm using a recursive formulation of the K-order Chebyshev polynomials to approximate the filters. Kipf et al.\cite{b10} proposed a graph convolutional network (GCN) with a faster localized graph convolutional operation, which is the first-order approximation of Chebyshev polynomials, that is, $K=1$. 
Veli et al.\cite{b25} proposed a graph attention network (GAT), which stacking layers in nodes that are able to attend over their neighborhoods’ features, specifying different weights to different nodes in a neighborhood, without requiring costly matrix operation or depending on knowing the graph structure upfront.

Bianchi et al.\cite{b73} proposed a graph convolutional layer that provides a flexible frequency response, which is more robust to noise, and better captures the global graph structure. Bouritsas et al.\cite{b74} proposed a graph substructure network that is more expressive than Weisfeiler-Leman graph isomorphism test, which allows the model retains multiple attractive properties of standard GNNs, while being able to eliminate even hard instances of graph isomorphism. Ciano et al.\cite{b75} proposed a mixed inductive–transductive GNN model, study its properties and introduce an experimental strategy that help to understand and distinguish the role of inductive and transductive learning. Tiezzi et al. \cite{b76} proposed an approach to learning in GNNs based on constrained optimization in the Lagrangian framework. Learning both the transition function and the node states is the outcome of a joint process, in which the state convergence procedure is implicitly expressed by a constraint satisfaction mechanism, avoiding iterative epoch-wise procedures and the network unfolding.

However, in EEG emotion recognition, some GNN-based methods\cite{b11} only consider second-order or third-order neighbors to avoid over-smoothing, which may result in the loss of valuable information between distant nodes. Thus, the spatial jigsaw puzzle was applied to challenge this problem.

\subsection{Multi-Task Learning}
Multi-task learning is an effective machine learning method and has shown its advantages in many fields, including computer vision\cite{b50}\cite{b21}, natural language processing\cite{b51}, and speech recognition\cite{b52}.
Ruder et al.\cite{b49} introduced two commonly used multi-task learning methods in deep learning, clarifying the working principle of multi-task learning as well as pointing out that properly designed pretext tasks can encourage the model to learn a more general representation while decreasing the risk of overfitting.
Compared with a single task, multi-task learning combines multiple related tasks and utilizes all the data from each task so that the knowledge on each task is shared. 
Additional information on the associated tasks is also obtained in multi-task learning models, resulting in significant improvements in the learning ability, generalization capability, and robustness of the model\cite{b17}. 
However, considering the different significance of each task, the weight of each task should be dynamic.
Sener et al.\cite{b20} regarded multi-task learning as a multi-objective optimization problem and proved that optimizing the upper bound of the multi-objective loss can obtain the Pareto optimal solution. 
Kendall et al.\cite{b21} proposed a principled approach to multi-task deep learning that weighs multiple loss functions by considering the homoscedastic uncertainty of each task to avoid the cost of manual tuning.
Benefiting from these advantages, in this work, multi-task learning framework is adopted to learn more generalization features and reduce the risk of overfitting.

\begin{figure*}[htb] 
	\centering 
	\includegraphics[width=0.95\textwidth]{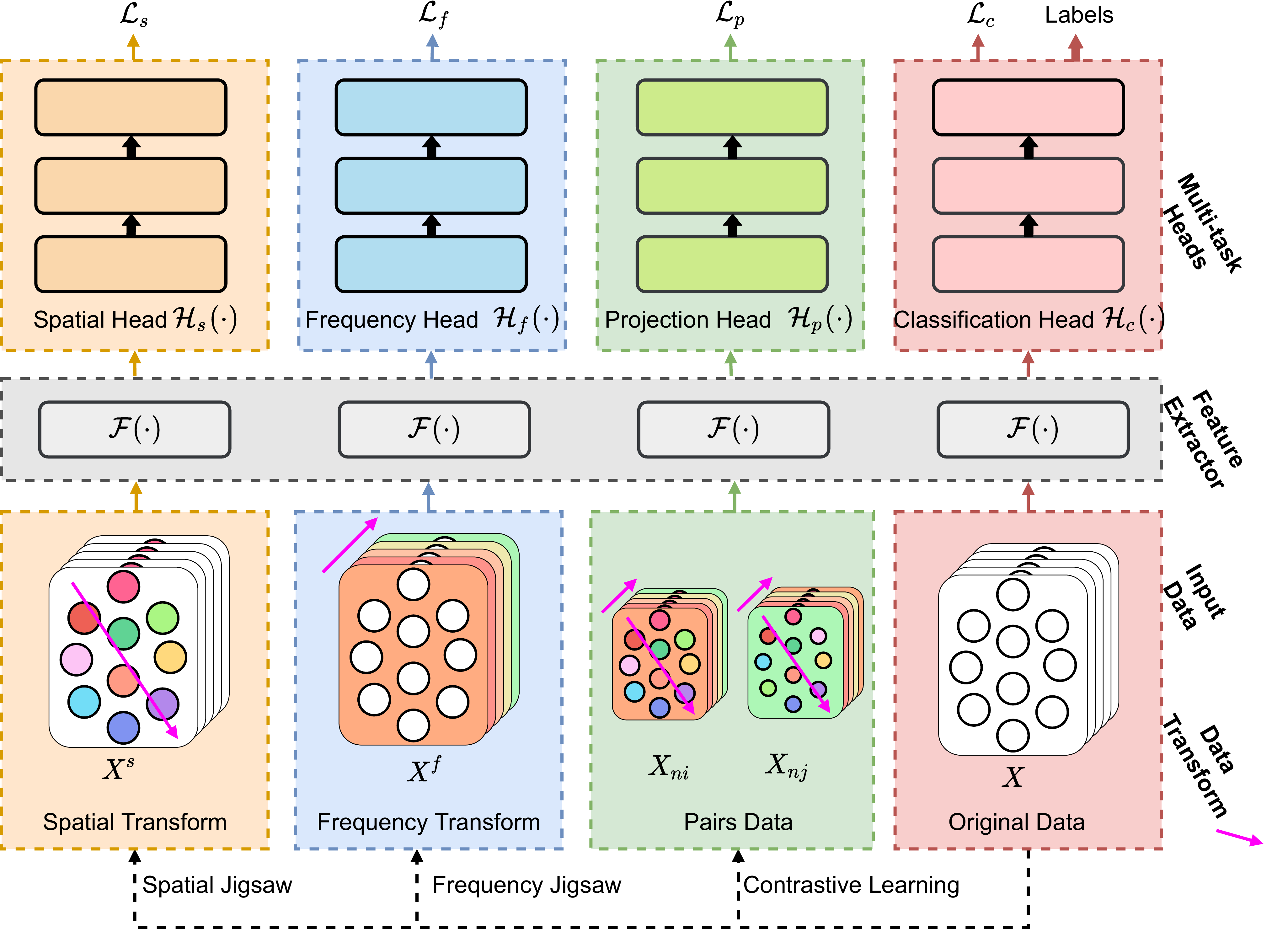}
	\caption{Framework of GMSS. In the unsupervised training mode, for the upstream task, the original graph data are not used to train the network. For the downstream task, the feature extractor is frozen and only the pink part with a substituted one linear layer is executed. In the supervised training mode, all the parts are executed simultaneously.} 
	\label{Fig.main2} 
\end{figure*}

\subsection{Self-Supervised Learning}
Self-supervised learning is a popular method for learning intrinsic information using unlabeled data\cite{b26}. Generally, self-supervised learning applies the attributes of data to generate pseudo labels as opposed to human-annotated labels to train the network.
Based on the different data attributes used in the design, there are four categories of pretext tasks: generation-based, context-based, free semantic label-based, and cross-modal-based\cite{b26}.
In the visual feature learning field, 
context-based pretext tasks mainly employs spatial structure, temporal structure, and context similarity for the design.
Many studies learn the general features of images by predicting the relative position of the patches to solve jigsaw puzzle tasks, thereby solving the problem of image classification \cite{b28}\cite{b29}\cite{b30}. Gidaris et al.\cite{b31} applied a 2D rotation to the image to construct the pretext task and then predicted the rotation angle to enable the model to learn the position, type, and posture of objects in the image. Carno et al. \cite{b32} used a clustering method to generate pseudo labels for images and combined learning neural network parameters and result features to obtain more abundant semantic information. 
Mathilde et al.\cite{b68} proposed a method for unsupervised learning of visual features by contrasting cluster assignments (SwAV), which takes advantage of contrastive methods without requiring to compute pairwise comparisons.
He et al.\cite{b53} suggested that momentum contrast (MoCo) would significantly narrow the gap between unsupervised representation learning and supervised representation learning. The performance of the contrastive SimCLR framework proposed by Chen et al.\cite{b19} on ImageNet surpasses that of supervised learning based models. Xinlei et al.\cite{b54} proposed a simple Siamese (SimSiam) network that achieved the best results without negative samples, large batches, and momentum encoders. In addition, the contrastive learning method was applied in the field of video processing, and achieved excellent performance at the time it was proposed\cite{b34}\cite{b35}. 
In the field of EEG emotion recognition, Xie et al.\cite{b72} proposed an innovative solution which contains six different transformations to learn high-level EEG representation (SSL-EEG). Mohsenvand et al.\cite{b71} present a framework for learning representations from EEG signals via contrastive learning which recombines channels from multi-channel recordings and trains a channel-wised feature extractor to learn EEG emotion representation (SeqCLR).
Inspired by self-supervised learning, in this work, two jigsaw puzzle tasks and a contrastive learning task were designed to assist with the learning of general EEG emotional features while circumventing the problem of EEG emotion noise labels.
Further, DeepCluster is a method based on clustering, and SwAV use a swapped prediction mechanism to predict the cluster assignment of a view from the representation of another view, SSL-EEG learn the EEG representations from complex signal transformation, while MoCo, SimCLR, SimSiam and SeqCLR are methods based on maximizing the similarity between positive pairs.
Compared with these methods above, our GMSS is a multi-task framework that incorporates multiple emotion-related tasks that utilizes all the data from each task so that the knowledge on each task is shared. This will helpful to obtain the additional information on the associated tasks that results in improving the learning ability, generalization capability, and robustness of the model. Another difference is that our self-supervised model concentrates on the characteristics of EEG emotion signal. For example, the jigsaw puzzle learning will force our model focus on the important brain regions and frequency bands of EEG signal, which are very important for emotion expression.

\begin{table}[H]
	\centering
	\caption{EEG electrodes associated with each brain region in the experiment.}\label{tab:1}
	\begin{tabular}{|c|c|}
		\hline
		\textbf{Brain region} & \textbf{Electrode name}                                                  \\ \hline
		Pre-Frontal           & AF3, FP1, FPZ, FP2, AF4                                                      \\ \hline
		Frontal               & F1, FZ, F2, FC1, FCZ, FC2                                                     \\ \hline
		Left Frontal          & F7, F5, F3, FT7, FC5, FC3                                                     \\ \hline
		Right Frontal         & F4, F6, F8, FC4, FC6, FT8                                                     \\ \hline
		Left Temporal         & T7, C5, C3, TP7, CP5, CP3                                                     \\ \hline
		Right Temporal        & C4, C6, T8, CP4, CP6, TP8                                                     \\ \hline
		Central               & \begin{tabular}[c]{@{}c@{}}C1, CZ, C2, CP1, CPZ,\\ CP2, P1, PZ, P2\end{tabular} \\ \hline
		Left Parietal         & P7, P5, P3, PO7, PO5, CB1                                                     \\ \hline
		Right parietal        & P4, P6, P8, PO6, PO8, CB2                                                     \\ \hline
		Occipital             & PO3, POZ, PO4, O1, OZ, O2                                                     \\ \hline
	\end{tabular}
\end{table}

\section{Graph-Based Multi-Task Self-Supervised Learning for EEG Emotion Recognition}
The goal of GMSS is to capture general and discriminative EEG emotion features using multi-task self-supervised learning, as illustrate in Fig. \ref{Fig.main2}. Three self-supervised tasks are designed to achieve this goal under unsupervised and supervised modes. 
These tasks share a common feature extractor. There are four task heads, i.e., Spatial Head $\mathcal{H}_s(\cdot)$, Frequency Head $\mathcal{H}_f(\cdot)$, Projection Head $\mathcal{H}_p(\cdot)$, Classification Head $\mathcal{H}_c(\cdot)$. $\mathcal{H}_s$ and $\mathcal{H}_f$ are employed for spatial puzzle and frequency puzzle respectively. $\mathcal{H}_p$ is adopted to project the learned representation into feature space. $\mathcal{H}_c$ is used for emotion recognition. Each head consists of three fully connected layers.

\begin{figure}[htb] 
	\centering 
	\includegraphics[width=.49\textwidth]{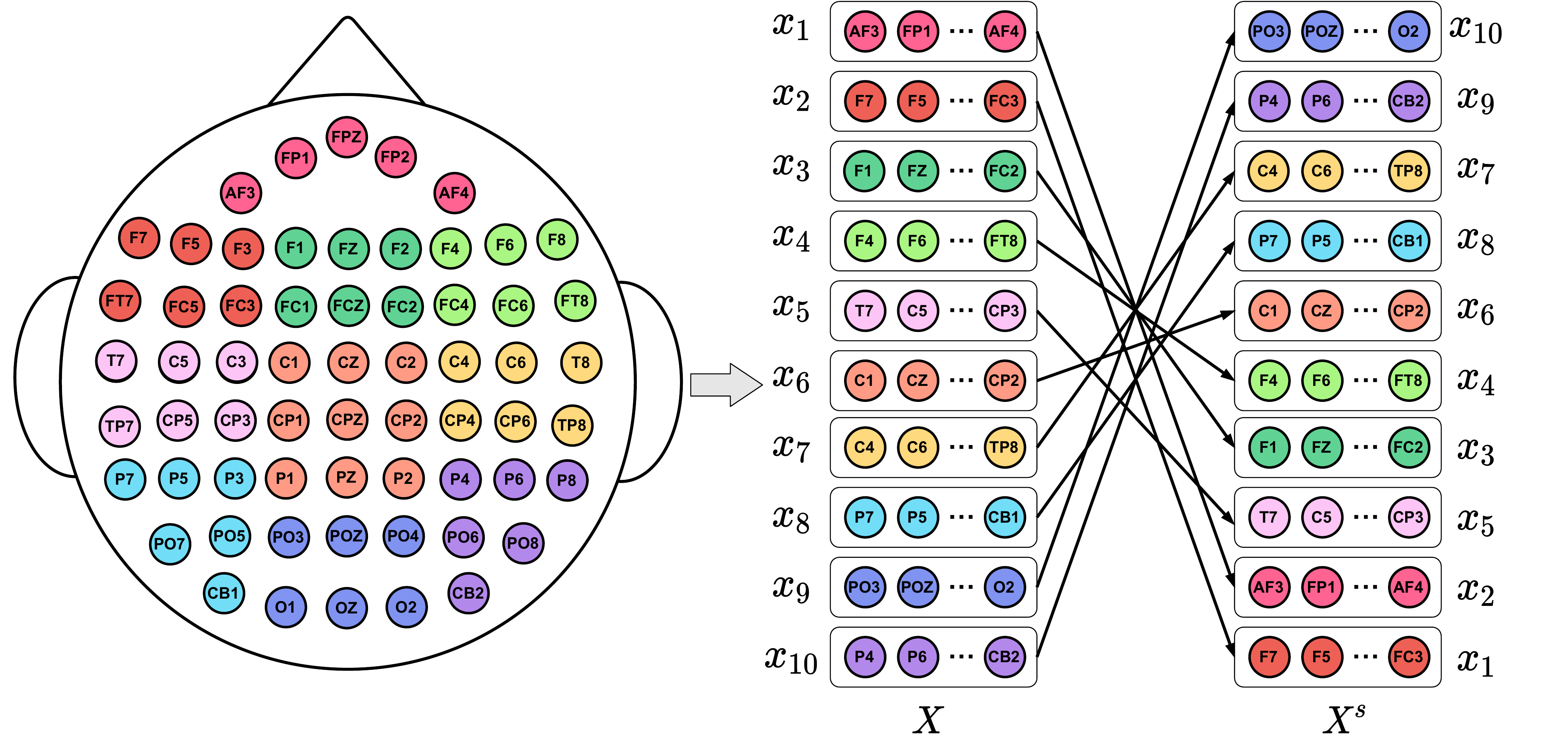}
	\caption{Spatial jigsaw puzzle. The 62 electrodes are divided into 10 blocks according to the location of brain regions. The placement of these channels is relocated while keeping the original connection based on the topology of the scalp. The spatial jigsaw puzzle task is to identify which of the 128 classes the channels reorganized by blocks belong.}
	\label{Fig.main3} 
\end{figure}

\subsection{Multiple Self-Supervised Tasks}
To learn more generalized and discriminative features and alleviate the noise problem of EEG emotion labels, multiple self-supervised learning tasks are considered, including the spatial jigsaw puzzle task, the frequency jigsaw puzzle task, and the contrastive learning task. Each of these three pretext tasks is described in depth.

\subsubsection{Spatial Jigsaw Puzzle}
The spatial jigsaw puzzle aims to capture the spatial patterns of EEG electrodes in different brain regions.
Due to the different effects of brain regions on emotion expression, the spatial jigsaw puzzle task is defined as a series of brain region permutations\cite{b36}\cite{b37}\cite{b38}\cite{b48}. 
As shown in Table \ref{tab:1}, the original EEG data $X \in \mathbb{R}^{n \times d}$ are partitioned into 10 blocks according to the location of the brain regions, denoted as $X=(\widetilde{X}_1, \widetilde{X}_2, \cdots, \widetilde{X}_{10})^\mathsf{T}$ where $\widetilde{X}_i\in\mathbb{R}^{n_i \times d}$, $\sum_{i=1}^{10} n_i =n, n_i>0$. Then, all brain region permutations can be obtained:

\begin{equation}\label{Eq1}
	\begin{cases}
		\hat{X}_{1} &= (\widetilde{X}_1, \widetilde{X}_2, \cdots, \widetilde{X}_{10} | y_1),  \\
		\hat{X}_{2} &=  (\widetilde{X}_1, \widetilde{X}_2, \cdots, \widetilde{X}_{9} | y_2), \\
		&\vdots  \\
		\hat{X}_{10!} &=  (\widetilde{X}_{10}, \widetilde{X}_{9}, \cdots, \widetilde{X}_{1} | y_{10!}),\\
	\end{cases}
\end{equation}

\noindent where $\hat{X_i}$ and $y_i$ represent the i-th permutation and its serial number, respectively. There are $10!=3628800$ permutations in total. 
The goal is to distinguish which permutation the spatial transformed data corresponds to.
However, it is quite challenging to distinguish these massive permutations for self-supervised pretext tasks. 
Therefore, we develop a $R_k(\cdot)$ operator. $R_k(\cdot)$ selects the $k$ permutations with maximum Hamming distance from the full permutation of Eq. (\ref{Eq1}) and randomly transformed the input data to one of the $k$ permutations. We define a unique pseudo label for each of these $k$ permutations, generating $k$ different kinds of pseudo labels in total, with a range from 1 to $k$. Each input data is randomly transformed into one of the $k$ permutations and the corresponding unique pseudo labels are obtained. $k$ is set to 128. The overall permutation is displayed in Fig. \ref{Fig.main3}, and is formulated as follows:
\begin{equation}\label{}
	({X}^s, {y}^s) = R_{128}(X),
\end{equation}
\noindent where ${X}^s$ is the generated EEG data with pseudo label $y^s\in\mathbb{Z}^{128}_+$.

To recognize these spatial jigsaw puzzles, a classification head $\mathcal{H}_s(\cdot)$ is applied, and cross entropy is adopted as the loss function. Formally, the loss of spatial jigsaw puzzle tasks can be expressed as $\mathcal{L}_s$:
\begin{equation}\label{}
	\mathcal{L}_s = -\sum_{i=1}^{N} \bar{y}_i^s  log(\mathcal{H}_s(\mathcal{F}({X}^{s}_{i}))) ,
\end{equation}
\noindent where $\mathcal{F}(\cdot)$ is the shared feature extractor, $\bar{y}_i^s$ is the one-hot encoding of the corresponding pseudo label $y_i^s$, and N is the number of training samples.

\begin{figure}[htb] 
	\centering 
	\includegraphics[width=.49\textwidth]{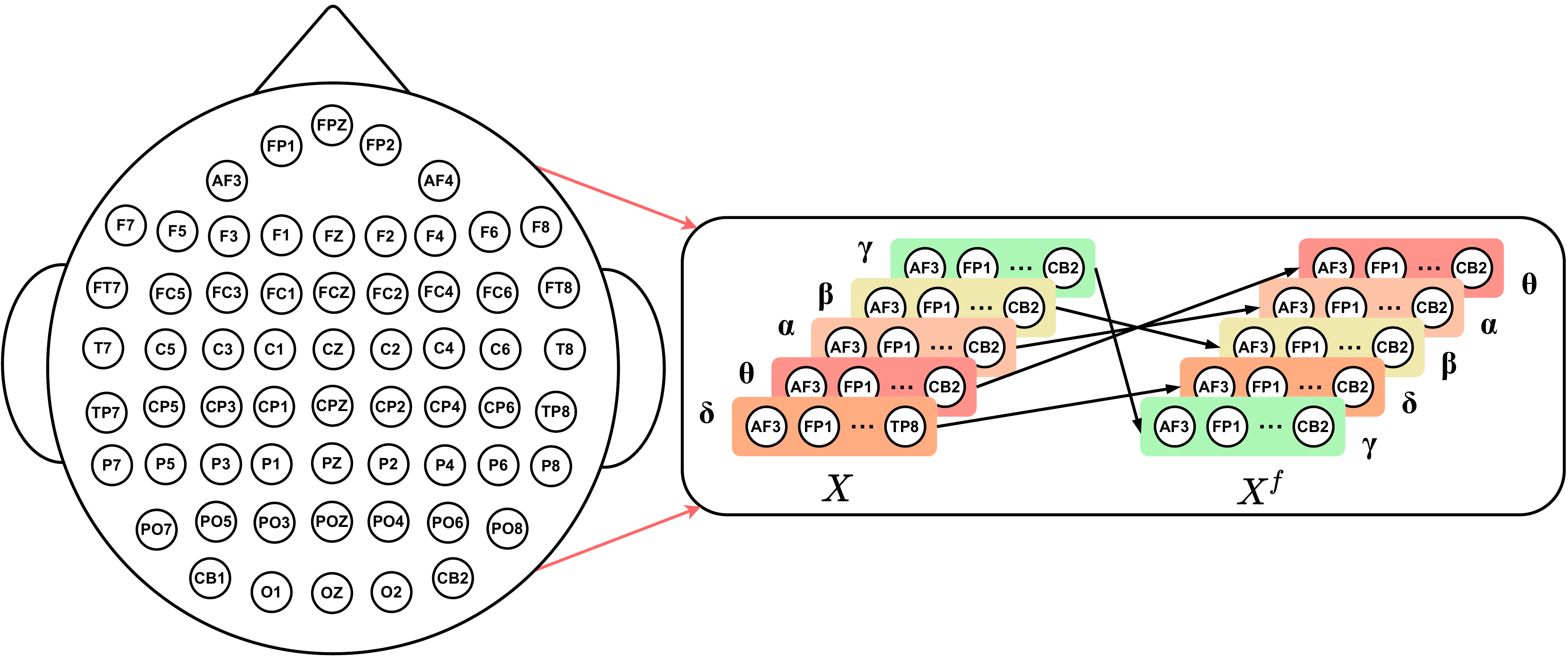}
	\caption{Frequency jigsaw puzzle. The frequency jigsaw puzzle transforms the frequency bands of each channel of an EEG emotion data in the same way. The goal of the frequency jigsaw puzzle is to figure out which of the 120 classes the scrambled EEG emotion data belong.} 
	\label{Fig.main4} 
\end{figure}

\subsubsection{Frequency Jigsaw Puzzle}
The frequency jigsaw puzzle task is designed to learn the inner relationship between frequency bands, explore the crucial frequency bands for EEG emotion recognition and improve the discrimination ability of the model. In general, as illustrated in Fig. \ref{Fig.main4}, the energy features of the EEG data are extracted from five emotion expression-related frequency bands, including $\delta$ (1-3 Hz), $\theta$ (4-7 Hz), $\alpha$ (8-13 Hz), $\beta$ (14-30 Hz), $\gamma$ (31-50 Hz).
Similar to the spatial jigsaw puzzle, the original EEG data $X$ are divided into five blocks according to different frequency bands, denoted as $(x_1, x_2, \cdots, x_{5})$, where $x_{j}\in\mathbb{R}^{n \times 1}$. 
The goal is to identify the corresponding permutation of the frequency transformed data. All frequency bands permutations can be obtained:

\begin{equation}
	\begin{cases}
		X_{1}^{'} &= (x_1, x_2, \cdots, x_{5} | y_1^{'}),  \\
		X_{2}^{'} &=  (x_1, x_2, \cdots, x_{4} | y_2^{'}), \\
		&\vdots  \\
		X_{5!}^{'} &=  (x_{5}, x_{4}, \cdots, x_{1} | y_{5!}^{'}),\\
	\end{cases}
\end{equation}
where $X_{j}^{'}$ and $y_j^{'}$ represent the j-th permutation and its serial number, respectively.
In the frequency jigsaw puzzle, the operator $R_k(\cdot)$ is applied to generate transformed data with pseudo label, and $k=120$:
\begin{equation}\label{}
	({X}^f, {y}^f) = R_{120}(X),
\end{equation}
\noindent where ${X}^f$ is the generated EEG data with pseudo label $y^f\in\mathbb{Z}^{120}_+$. 

To recognize these frequency jigsaw puzzles, a classification head $\mathcal{H}_f(\cdot)$ is applied and cross entropy is adopted as the loss function. Formally, the loss in the frequency jigsaw puzzle task can be expressed as follows:
\begin{equation}\label{}
	\mathcal{L}_f = -\sum_{j=1}^{N} \bar{y}_j^f  log(\mathcal{H}_f(\mathcal{F}({X}^{f}_{j}))) ,
\end{equation}
\noindent where $\mathcal{F}(\cdot)$ is the shared feature extractor and $\bar{y}_j^f$ is the one-hot encoding of the corresponding pseudo label $y_j^f$.

\begin{figure}[htb] 
	\centering 
	\includegraphics[width=.35\textwidth]{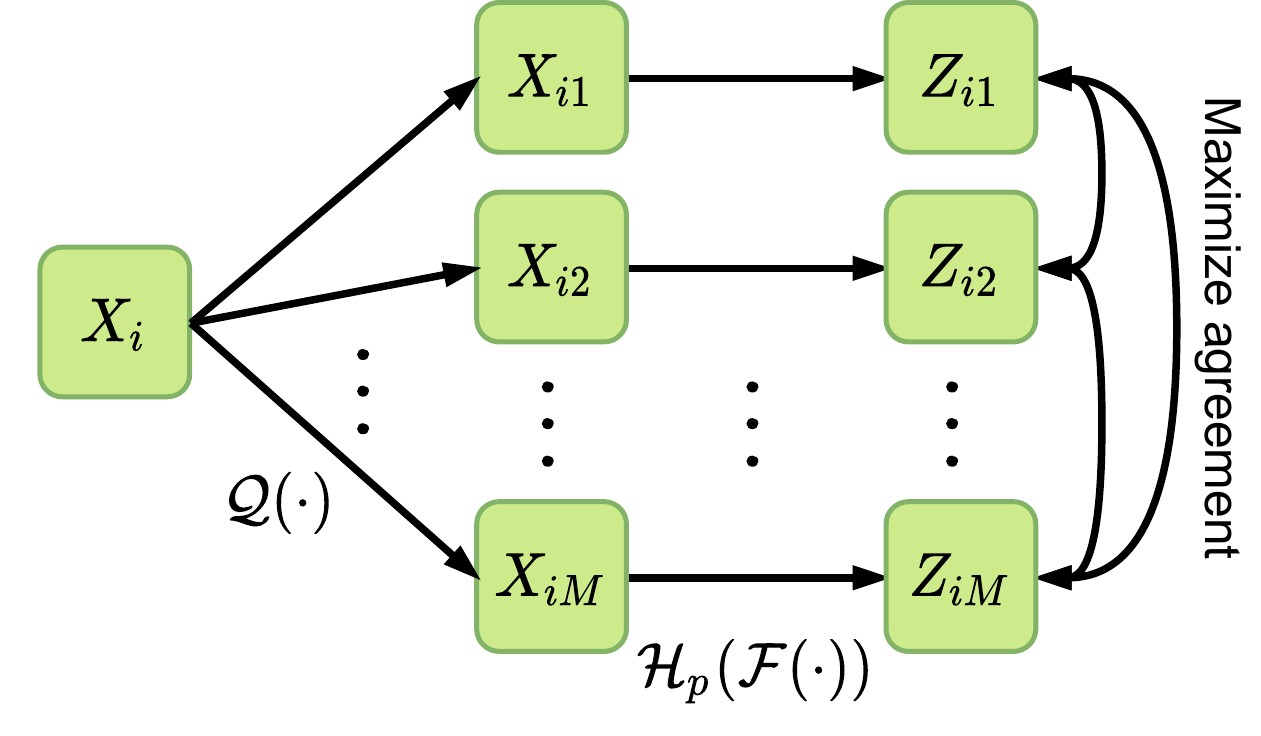}
	\caption{Contrastive learning. The original data applied with spatial transformation and frequency transformation to generate the pairs data.} 
	\label{Fig.main15} 
\end{figure}

\subsubsection{Contrastive Learning}
To further regularize feature learning and encourage the network to learn inherent representations, contrastive learning is adopted to map the transformed data into a common feature space.
The purpose is to maximize the agreement between the different augmented data of the same EEG emotion data, as shown in Fig. \ref{Fig.main15}. To ensure that positive pairs move closer and negative pairs move far away in feature space, a data augmentation operation $\mathcal{Q}(\cdot)$ is defined to consider the spatial and frequency transformations of the same original EEG emotion data. 
For each original EEG emotion data $X_i, i\in\{1, 2, \cdots, N\}$, M augmented data $\{X_{i1}, X_{i2}, \cdots, X_{iM}\} = \mathcal{Q}(X_i)$ are obtained. As a result, each augmented data has $(M-1)$ positive pairs and $(N-1)\times M$ negative pairs. In total, $N \times M$ augmented data are obtained by:

\begin{equation}
	\begin{aligned}
		\{X_{nm} ; n \in \{1,2,\cdots,N\}, m \in \{1,2,\cdots,M\}\} = \\
		\mathcal{Q}(X_1)\cup\mathcal{Q}(X_2)\cup\cdots\cup\mathcal{Q}(X_N),
	\end{aligned}	
\end{equation}

\noindent where ${X_{nm} \in \mathbb{R}^{n\times d}}$ is the m-th transformation of the n-th EEG sample.

Similar to SimCLR\cite{b19}, a projection head $\mathcal{H}_p(\cdot)$ is applied to map the EEG emotion data onto the feature space, that is, $ {Z}_{nm} = \mathcal{H}_p(\mathcal{F}({X}_{nm})) $. 
The similarity of two data points is quantitatively described by the dot product, which normalizes u and v through the $\ell_{2}$-norm.
i.e., $ sim(u, v) = u^\mathsf{T}v/\left\|u\right\|\left\|v\right\| $. 
Then, the loss of all positive pairs $\ell_{n}$ of sample $X_n$ is calculated as follows:
\begin{equation}
	\ell_{n} = -log
	\frac{g_{+}}{g_{+} + g_{-}}	,
\end{equation}

\begin{equation}	
	g_{+} = \sum_{i=1}^{M-1} \sum_{j=i+1}^{M}  exp(sim({Z}_{ni}, {Z}_{nj})/\tau) 	,	
\end{equation}

\begin{equation}\label{eq1}
	g_{-} = \sum_{o=1}^{M} \sum_{t=1}^{N} \sum_{w=1}^{M} exp(sim({Z}_{no}, {Z}_{tw})/\tau), t \neq n,
\end{equation}

\noindent where $({Z}_{ni}, {Z}_{nj})$ are positive pairs, and $({Z}_{no}, {Z}_{tw})$ are negative pairs. $\tau$ is the temperature parameter and is set to 0.5. Furthermore, the arithmetic average of the loss of all positive pairs' $\ell_{n}$ of all samples is calculated for backpropagation as follows:

\begin{equation}
	\mathcal{L}_{p} = \frac{1}{N} \sum_{n=1}^{N} \ell_{n},
\end{equation}

\subsection{Training Mode for EEG Emotion Recognition}
Two modes of training are provided: unsupervised and supervised. The feature extractor $\mathcal{F(\cdot)}$ in both modes are the same. When training the feature extractor, the distinction is in the presence or absence of ground-truth emotion labels. In the unsupervised mode, instead of using ground-truth emotion labels, the feature extractor $\mathcal{F(\cdot)}$ is trained only on the self-supervised tasks mentioned above.
Then, the frozen feature extractor $\mathcal{F(\cdot)}$ is transferred to the downstream task and the performance is verified using a linear classifier. In the supervised mode, a joint training strategy is adopted. The network is simultaneously trained on self-supervised tasks and supervised tasks. 
To avoid manually tuning the weights of the different loss functions, the total loss function is defined by considering the homoscedastic uncertainty of each task \cite{b21}. In particular, the training loss $\mathcal{L}$ is calculated as follows:
\begin{equation}
	\begin{aligned}
		\mathcal{L} = 
		\frac{1}{\sigma_{\mathcal{L}_{s}}^2} \mathcal{L}_{s} &+ \frac{1}{\sigma_{\mathcal{L}_{f}}^2} \mathcal{L}_{f} + \frac{1}{2\sigma_{\mathcal{L}_{p}}^2} \mathcal{L}_{p} + log(\sigma_{\mathcal{L}_{s}}\sigma_{\mathcal{L}_{f}}\sigma_{\mathcal{L}_{p}}) \\&+
		\psi\cdot(\frac{1}{\sigma_{\mathcal{L}_{c}}^2} \mathcal{L}_{c} + log(\sigma_{\mathcal{L}_{c}})), 
	\end{aligned}	
\end{equation}

\begin{equation}
	\psi = 
	\begin{cases}
		0, \quad unsupervised\quad mode,\\
		1, \quad supervised\quad mode,
	\end{cases}
\end{equation}
\noindent where $\mathcal{L}_{c}$ is the cross entropy loss of supervised EEG emotion classification task; $\sigma_{\mathcal{L}_{s}}, \sigma_{\mathcal{L}_{f}}, \sigma_{\mathcal{L}_{p}}$ and $\sigma_{\mathcal{L}_{c}}$ are the observation noise scalars of the corresponding tasks\cite{b21}. $\psi$ is the mode-selection operator. The observation noise scalar $\sigma$ is a principled approach to multi-task deep learning which weighs multiple loss functions by the homoscedastic uncertainty of each task. This allows us to simultaneously learn various quantities with different units or scales in both classification and regression settings, which can balance these weightings optimally, resulting in superior performance. These scalars can be calculated as learnable parameters which change constantly during the model training process and the initial values are 1.

\begin{figure}[H] 
	\centering 
	\includegraphics[width=.3\textwidth]{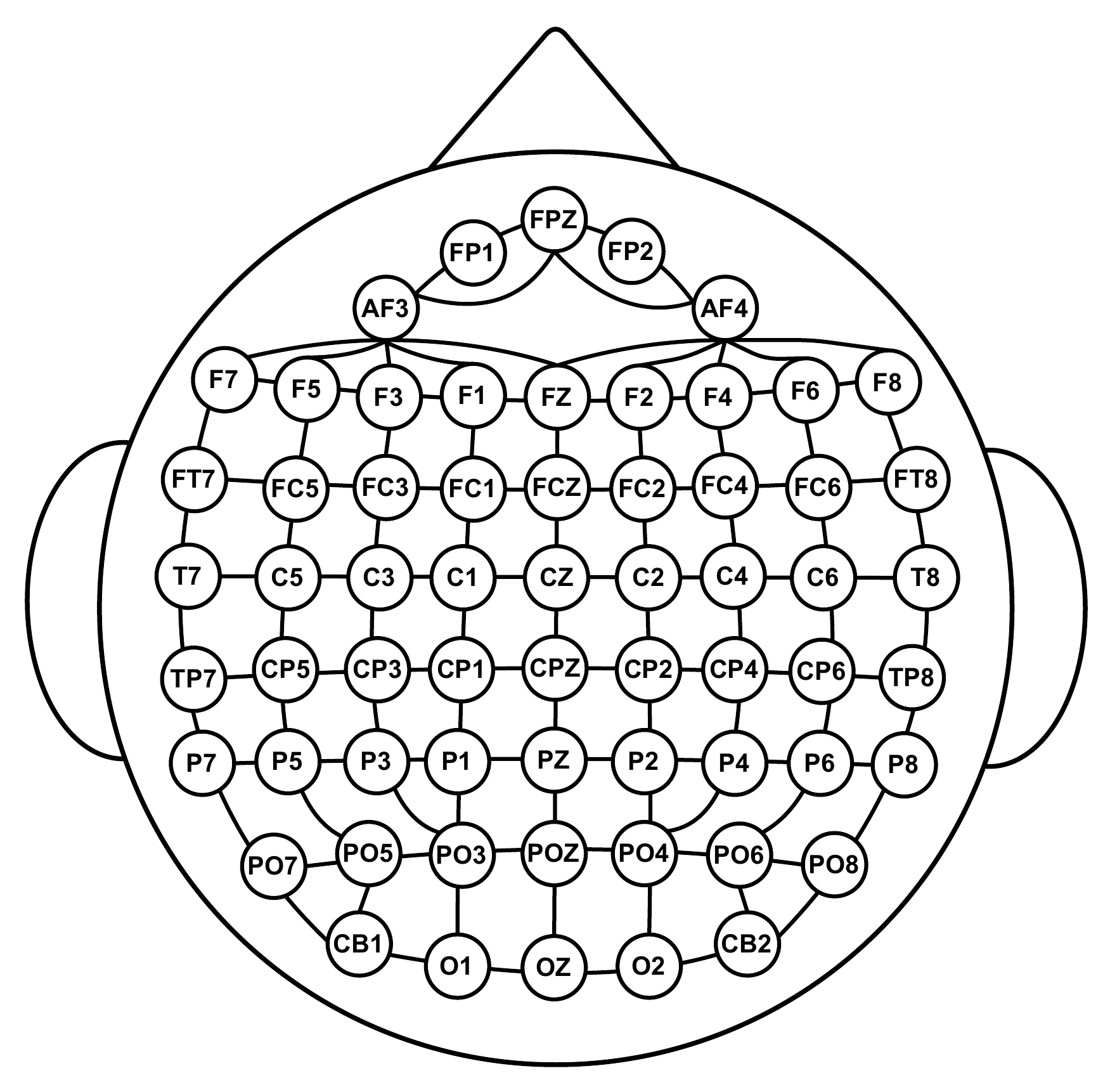} 
	\caption{EEG graph structure and adjacency matrix $A$ construction.} 
	\label{Fig.main1} 
\end{figure}

\subsection{Feature Extractor of GMSS}
As shown in Fig. \ref{Fig.main1}, an undirected graph $\mathcal{G(V, E)}$ is employed to model the EEG data. Meanwhile, the adjacency matrix $A$ of the EEG data was obtained. In $\mathcal{G(V, E)}$, $\mathcal{V}$ denotes the set of nodes where $\mathcal{|V|}=n$; each node has $d$ dimensions. As a result, nodes can be represented by the feature matrix $X\in \mathbb{R}^{n \times d}$. $\mathcal{E}$ denotes a set of edges between nodes. $(v_i, v_j)$ is the edge between nodes $v_i$ and node $v_j$, that is, $(v_i, v_j) \in \mathcal{E}$. The adjacency matrix $A \in \mathbb{R}^{n \times n}$ contains the topological information of the undirected graph, that is, the EEG data. $D$ is the degree matrix of the vertices, and $L= D - A$ is the combinatorial Laplacian matrix. In this study, $n$ denotes the channel number of EEG data; $d$ denotes the number of frequency bands and $d=5$. The energy feature is extracted from five bands, namely, $\delta$ (1-3 Hz), $\theta$ (4-7 Hz), $\alpha$ (8-13 Hz), $\beta$ (14-30 Hz), $\gamma$ (31-50 Hz).

In the GMSS model, Chebyshev polynomials are employed instead of the convolution kernel of SCNN \cite{b42} in the spectral domain, so that there are only k parameters in the convolution kernel, and feature decomposition is not required, reducing the computational load. Thus, the feature extractor $\mathcal{F}(\cdot)$ of the GMSS can be formulated as: 
\begin{equation}\label{}
	\mathcal{F}(X) = \sigma(\sum_{k=0}^{K-1}\beta_{k}T_{k}(\widetilde{L})X) ,
\end{equation}
where $\sigma(\cdot)$ is the activation function; $X$ is the input EEG emotion data; $\beta_{k}$ refers to the learning parameters in network training; and $T_{k}(\cdot)$ is the Chebyshev polynomial of order $K$. Additionally, $\widetilde{L} = 2L/\lambda_{max}-I \label{eq2} $, where $\lambda_{max}$ is the maximum eigenvalue of Laplace matrix $L$.
In this study, we set $K=2$ to avoid over-smoothing.

\begin{table*}[htb]
	\centering
	\caption{Subject-dependent and subject-independent classification accuracy (mean/std) for unsupervised mode on SEED, SEED-IV, and MPED datasets}\label{tab:3}
	\begin{threeparttable}
		\begin{tabular}{ccccccc}
			\toprule
			\multirow{2}{*}{Model} & \multicolumn{2}{c}{SEED}        & \multicolumn{2}{c}{SEED-IV}      & \multicolumn{2}{c}{MPED}        \\
			& dependent & independent & dependent & independent & dependent & independent \\
			\midrule
			DeepCluster\cite{b44} & 74.60/12.17*      & 59.01/17.65*          & 49.60/10.28*     & 44.54/09.88*          & 26.38/05.59*     & 23.25/04.86*          \\
			
			MoCo\cite{b53} & 76.58/10.72*      & 58.26/15.05*          & 49.40/10.99*     & 46.19/10.04*         & 27.47/05.27*     & 23.86/04.66*          \\
			
			SwAV\cite{b68} & 77.81/10.15*    & 58.65/16.66*    & 52.03/14.71*    & 49.28/10.44*    & 27.91/05.05*    & 23.50/04.81*    \\
			
			SimCLR\cite{b19} &81.79/11.15*       & 63.45/15.96*          & 52.47/11.57*     & 50.07/11.17*          & 29.53/05.36*     & 24.21/05.10*          \\
			
			SimSiam\cite{b54} &80.18/10.53*       & 63.95/11.95*          & 53.71/11.98*     & 51.24/12.47*          & 28.19/05.88*     & 24.31/04.61*          \\
			SSL-EEG\cite{b72} &83.32/09.20*       & 67.52/12.73*          & 63.59/19.82*     & 53.62/08.47*          & 25.22/04.25*     & 21.87/02.53*          \\
			SeqCLR\cite{b71} &82.91/08.97*       & 64.56/11.89*          & 63.13/15.41*     & 50.75/07.71*          & 30.47/06.07*     & 23.33/03.89*          \\
			\midrule
			GMSS                   & \textbf{89.18/09.74 } & \textbf{76.04/11.91 } & \textbf{65.61/17.33 } & \textbf{62.13/08.33 } & \textbf{34.81/06.88 } & \textbf{26.97/05.01 } \\
			\bottomrule
		\end{tabular}
		
		\begin{tablenotes}
			\item[*] indicates the experiment results obtained by our own implementation.
			\item Note: For the subject-dependent experiment, we calculate the average accuracy based on the results of  all the sessions. While for the subject-independent experiment, we calculate the average accuracy based on the results of  all the subjects.
			
		\end{tablenotes}
		
	\end{threeparttable}
\end{table*}

\begin{table*}[htb]
	\centering
	\caption{Subject-dependent and subject-independent classification accuracy (mean/std) for supervised mode on SEED, SEED-IV, and MPED datasets}\label{tab:2}
	\begin{threeparttable}
		\begin{tabular}{ccccccc}
			\toprule
			\multirow{2}{*}{Model} & \multicolumn{2}{c}{SEED}        & \multicolumn{2}{c}{SEED-IV}      & \multicolumn{2}{c}{MPED}        \\
			& dependent & independent & dependent & independent & dependent & independent \\
			\midrule
			SVM\cite{b55}& 83.99/09.72& 56.73/16.29& 56.61/20.05& 37.99/12.52& 32.39/09.53 & 19.66/03.96\\
			DGCNN\cite{b11}& 90.40/08.49& 79.95/09.02& 69.88/16.29& 52.82/09.23& 32.37/06.08 & 25.12/04.20\\
			DANN\cite{b21}& 91.36/08.30& 75.08/11.18& 63.07/12.66& 47.59/10.01& 35.04/06.52 & 22.36/04.37\\
			BiDANN\cite{b46}& 92.38/07.04& 83.28/09.60& 70.29/12.63& 65.59/10.39& 37.71/06.04 & 25.86/04.92\\
			A-LSTM\cite{b41}& 88.61/10.16& 72.18/10.85& 69.50/15.65& 55.03/09.28&38.99/07.53& 24.06/04.58\\
			BiHDM\cite{b13}& 93.12/06.06&  85.40/07.53& 74.35/14.09& 69.03/08.66& 40.34/07.53& 28.27/04.99\\
			RGNN\cite{b58}& 94.24/05.95& 85.30/06.72& 79.37/10.54& 73.84/08.02&--- &---\\
			\midrule
			BiHDM w/o DA& 91.07/08.21&81.55/09.74& 72.22/14.69&67.47/08.22& 38.55/07.22& 27.43/04.96\\
			RGNN w/o DA& ---&81.92/09.35& ---&71.65/09.34& ---& ---\\
			\midrule
			GMSS & \textbf{96.48/04.63}&\textbf{86.52/06.22} & \textbf{86.37/11.45}&\textbf{73.48/07.41}& \textbf{40.16/06.08}& \textbf{28.49/04.42}\\
			\bottomrule
		\end{tabular}
		
		\begin{tablenotes}
			\item[---] indicates the experiment results are not reported on that dataset.
			\item Note: For the subject-dependent experiment, we calculate the average accuracy based on the results of  all the sessions. While for the subject-independent experiment, we calculate the average accuracy based on the results of  all the subjects.
		\end{tablenotes}

	\end{threeparttable}
\end{table*}

\section{Experiments}
In this section, experiments were conduct on the following three datasets to evaluate the performance of our model: SEED\cite{b40}, SEED-IV\cite{b39}, and MPED\cite{b41}. All three datasets were collected while subjects watched emotional video clips in a quiet, comfortable, and non-interfering environment. All three datasets were generated by recording EEG signals through the ESI NeuroScan system using 62 electrode channels positioned according to the 10-20 system\cite{b36}. These three datasets are introduced next along with the experimental results.

\subsection{Experimental Dataset}
\noindent \textbf{SEED}. In the SEED dataset, there are a total of 15 subjects. There are three sessions associated with each subject. In each session, there are a total of 15 film clips to induce happy, neutral, and sad emotions, and there are 5 film clips for each emotion. That is, there are 15 trials per session, and each trial has 185-238 samples, resulting in approximately 3400 samples per session.

\noindent \textbf{SEED-IV}. In the SEED-IV dataset, similar to SEED, there are 15 subjects, and three sessions for each subject. The difference is that each session includes four kinds of emotions: happy, neutral, sad, and fear. Each emotion has 6 different film clips. As a result, there are 24 trials, and each trial has 12-64 samples for each session. Consequently, each session has approximately 830 samples.

\noindent \textbf{MPED}. In the MPED dataset there are 30 subjects, and each subject has only one session. In a session, there are seven types of emotions: joy, funny, neutral, sad, fear, disgust, and anger. Each type of emotion has 4 related film clips. Therefore, there are 28 trials per session. Each trial consists of 120 samples and there are a total of 3360 samples in one session.

\subsection{Experimental Protocol}

To fully evaluate our model, two types of experiments are implemented: subject-dependent and subject-independent experiment. For the subject-dependent experiment, the training data and testing data are obtained from different EEG trials of the same subject. For the subject-independent experiment, the training data and testing data are obtained from different subjects.

For the subject-dependent experiment, the same experimental protocol is applied as in \cite{b7}\cite{b40}\cite{b13}\cite{b41}. That is, for the SEED dataset, the EEG data of the first nine trials are used in each session as training data and the remaining six trials in the session as testing data for each subject. For the SEED-IV dataset, the first sixteen trials of the session are used for each subject as training data and the remaining eight trials as testing data. For the MPED dataset, the EEG data of the first twenty-one trials in the session are adopted for the training data and the remaining seven trials in this session are the testing data for each subject.

For the subject-independent experiment, the leave-one-subject-out (LOSO) cross-validation strategy is used in \cite{b13}\cite{b43} for each subject. Namely, one subject’s EEG emotion data constituted the testing data, and the remaining subjects’ EEG emotion data constituted the training data. The process continued until all subjects' EEG emotion data are tested once.

\subsection{Experimental Details}
In the experiments, the released differential entropy (DE) in SEED and SEED-IV, and the short-time Fourier transform (STFT) in MPED are feed into the model as input. The size of the input X is $62\times 5$; the output dimensions of each electrode is 32; and $K=2$, that is, the graph convolution aggregated the information of the second-order neighbors. In particular, GMSS is implemented by pytorch on a Nvidia 3080 GPU. The model is trained using the Adam optimizer with a batch size of 100. The learning rate is 0.001, and the weight decay rate is 8e-5. 
The mean accuracy (ACC) and standard deviation (STD) are employed as evaluation criteria in all datasets.
The code of GMSS can be found at \url{https://github.com/CHEN-XDU/GMSS}.

\subsection{Experimental Results}

\begin{figure}[htbp]
	\centering
	\subfigure[SEED]{\includegraphics[width=0.48\columnwidth]{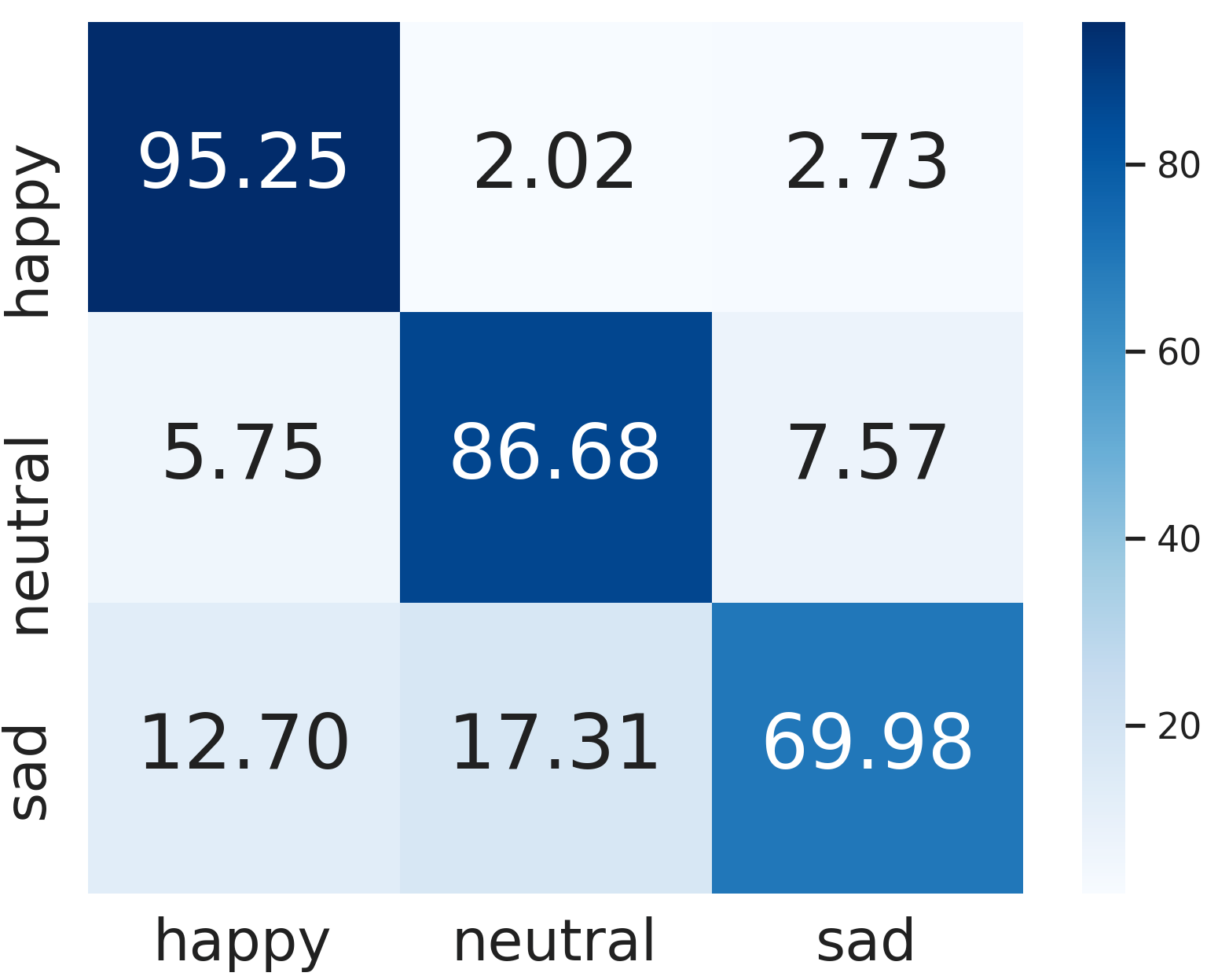}}
	\subfigure[SEED-IV]{\includegraphics[width=0.48\columnwidth]{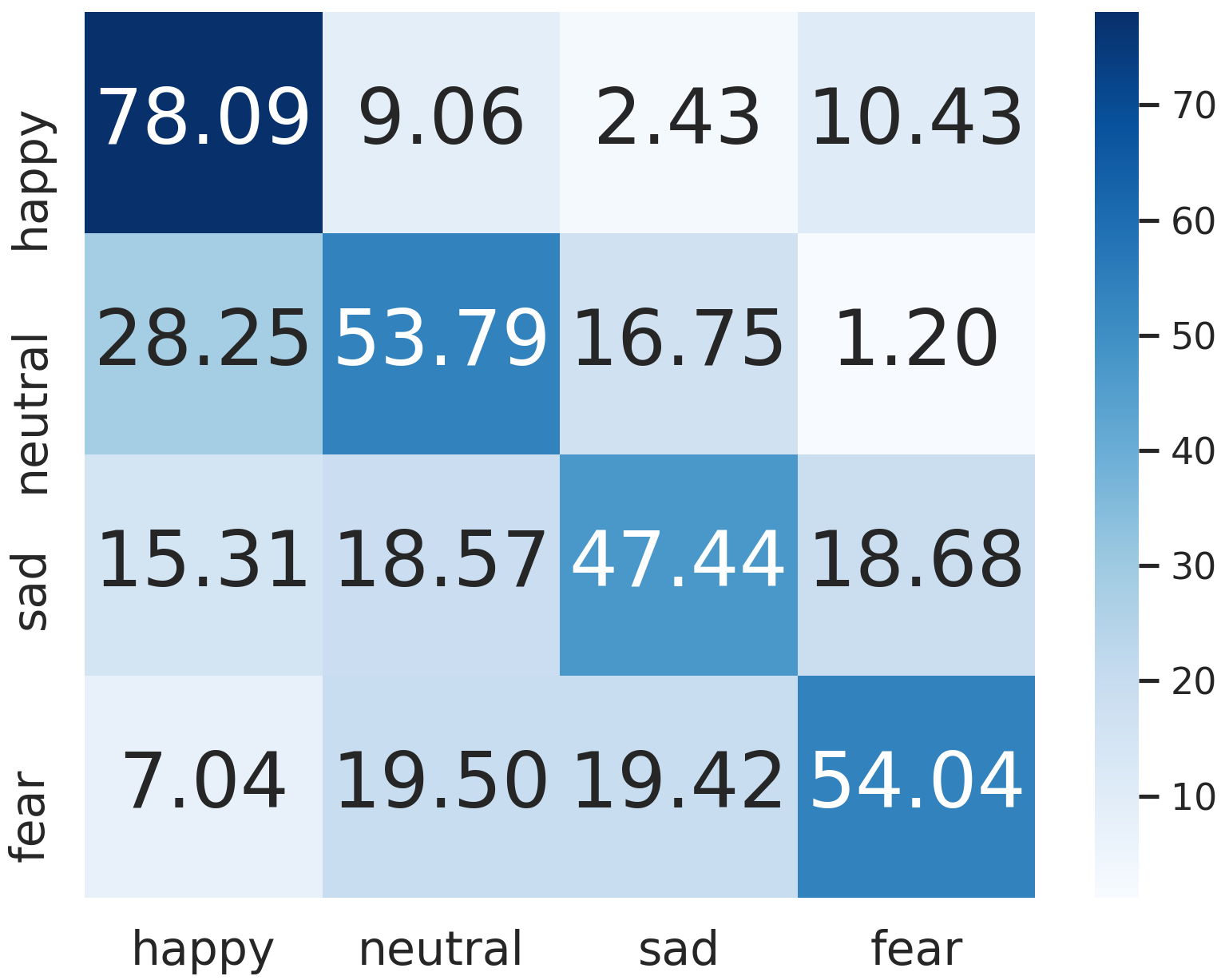}}
	\subfigure[MPED]{\includegraphics[width=0.95\columnwidth]{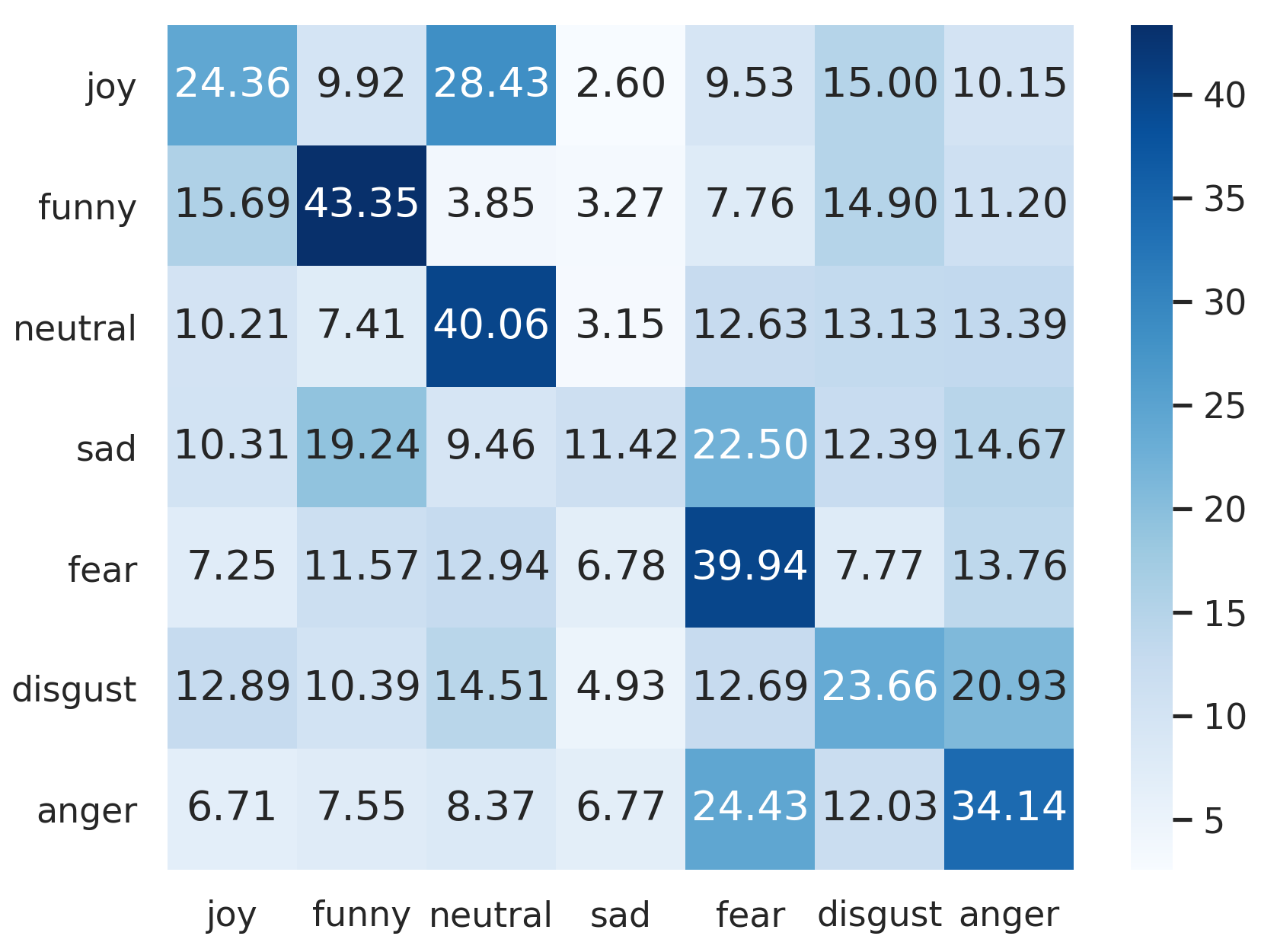}}
	\text{(1) Subject-dependent experimental results}
	\\
	
	\subfigure[SEED]{\includegraphics[width=0.48\columnwidth]{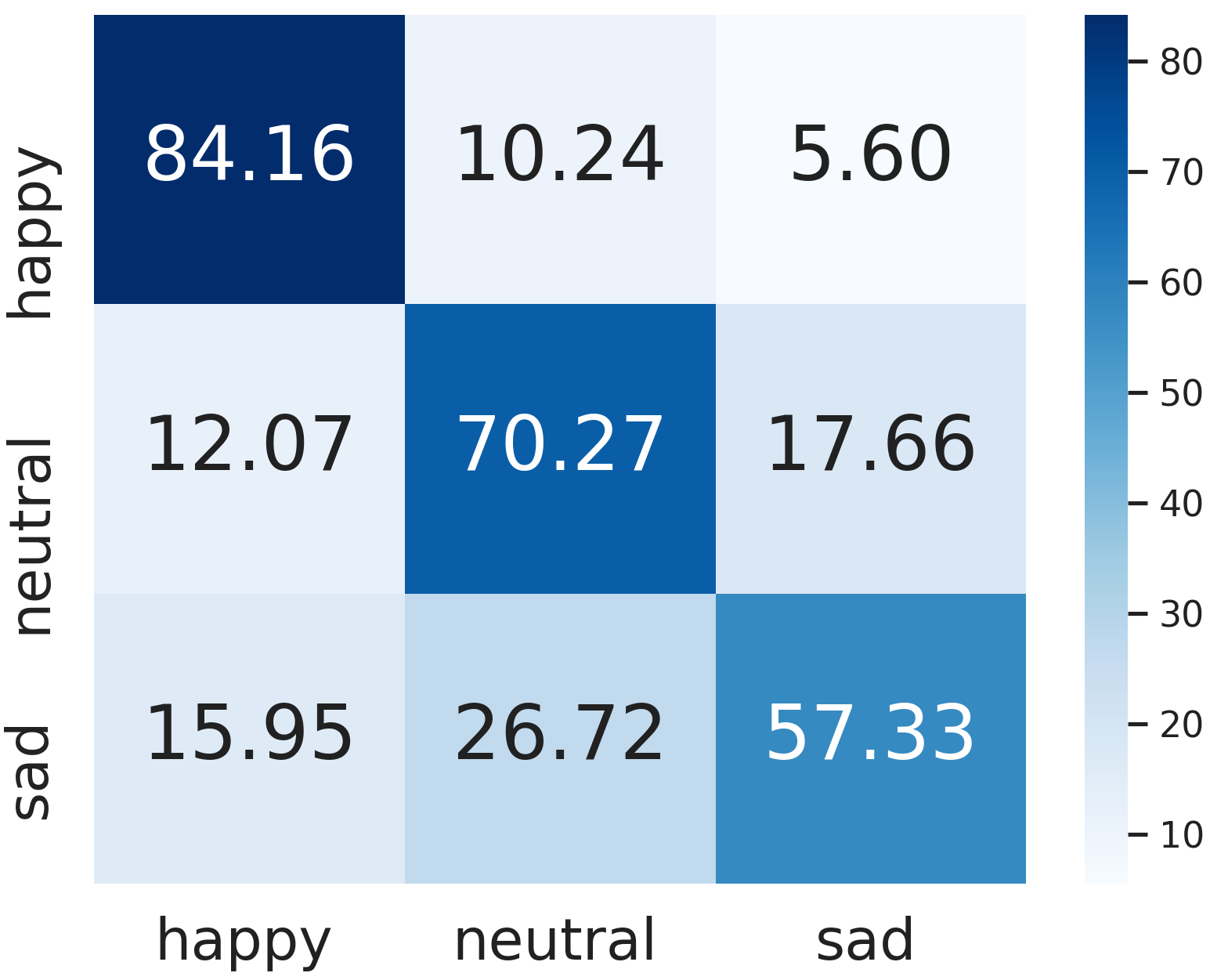}}
	\subfigure[SEED-IV]{\includegraphics[width=0.48\columnwidth]{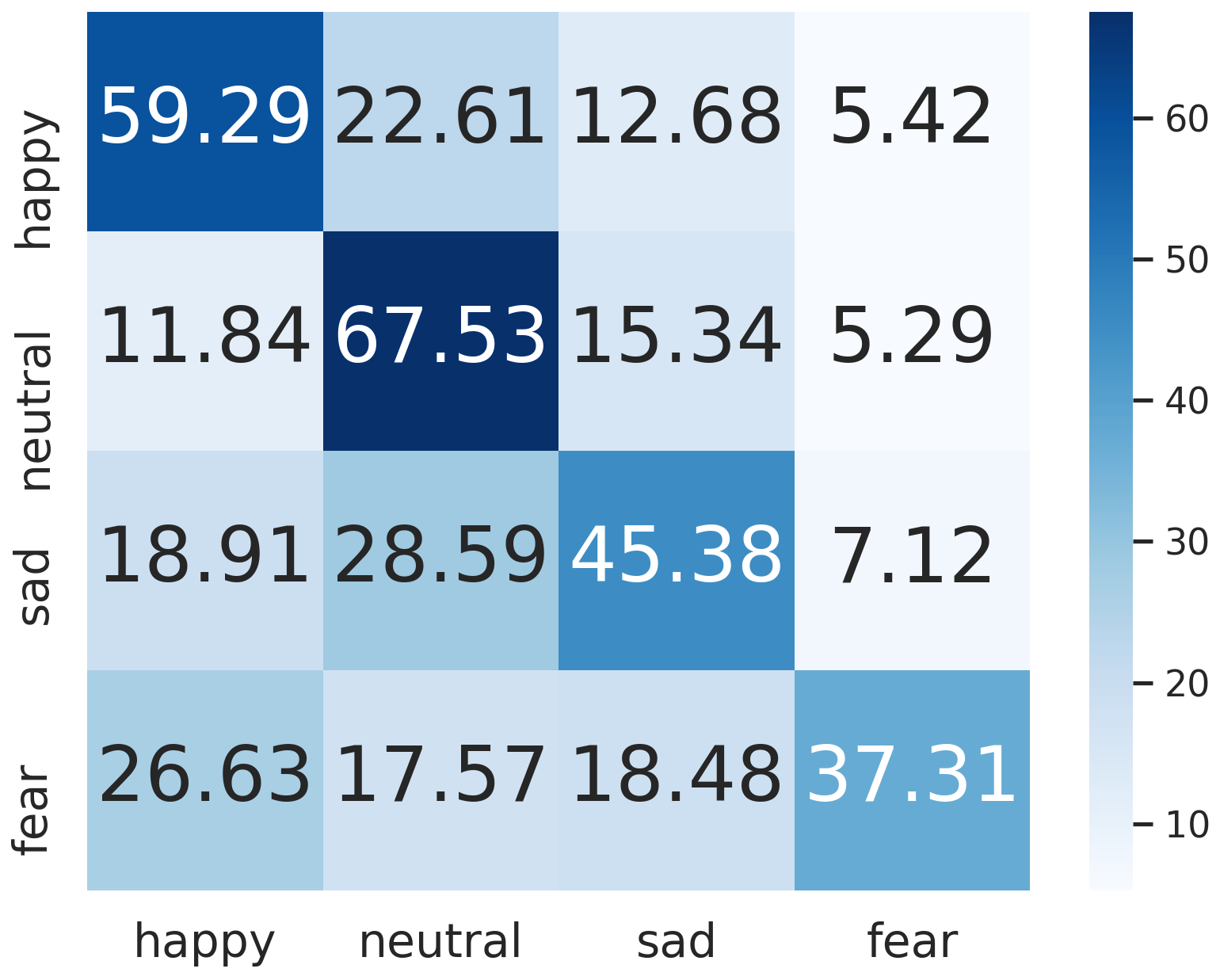}}
	\subfigure[MPED]{\includegraphics[width=0.95\columnwidth]{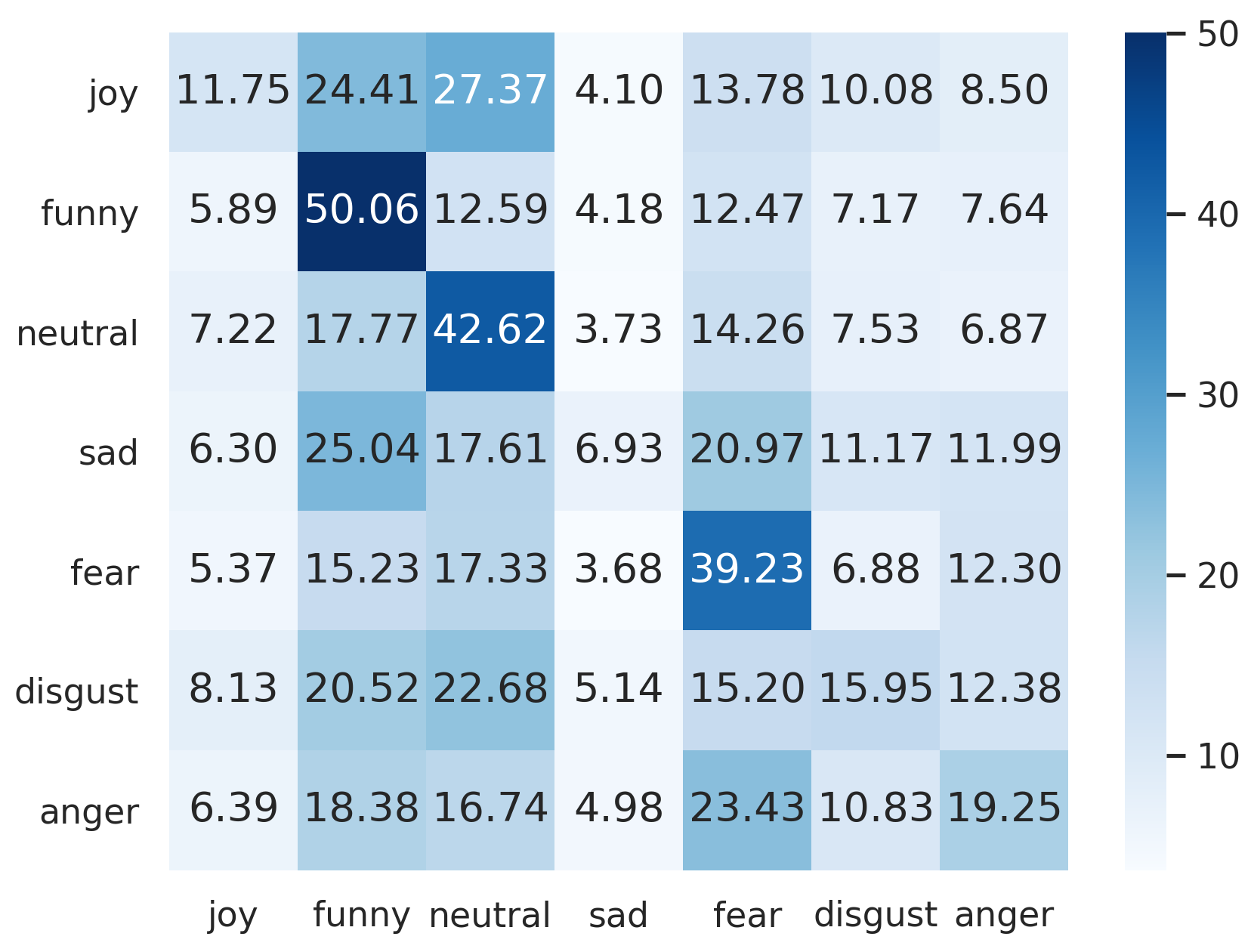}}
	\text{(2) Subject-independent experimental results}
	\\
	
	\caption{ Confusion matrices in unsupervised mode. (a)-(c) and (d)-(f) are the subject-dependent and subject-independent results on SEED, SEED-IV and MPED datasets, respectively.}
	\label{Fig.main6}
\end{figure}

\subsubsection{Unsupervised Mode}

In the upstream task, the model is trained by self-supervised pretext tasks, consisting of two jigsaw puzzle tasks and one contrastive learning task. In the downstream task, the frozen feature extractor is applied and a linear classifier is used to evaluate the performance of GMSS. 
We compared GMSS with two self-supervised EEG emotion recognition methods SSL-EEG\cite{b72} and SeqCLR\cite{b71}. In addition, since there are few methods based on self-supervised EEG emotion recognition and the code is not released, we also compared with some popular self-supervised methods in other fields such as DeepCluster\cite{b44}, MoCo\cite{b53}, SwAV\cite{b68}, SimCLR\cite{b19} and SimSiam\cite{b54}. These methods are reproduced and maintain the experimental protocol consistent with GMSS. 

For a fair comparison, all of these methods of other fields adopted the same feature extraction operation as GMSS. 
To fit the EEG emotion recognition task, MoCo, SwAV, SimCLR and SimSiam adopted the same data augmented as GMSS. 
The experimental results are shown in Table \ref{tab:3}. 
Concretely, GMSS improves the accuracy by 5.86\%, 8.52\%, 2.02\%, 8.51\%, 4.34\%, and 2.66\% compared with the existing SOTA methods in the subject-dependent and subject-independent experiments on SEED, SEED-IV, and MPED datasets, respectively.
Especially compared with MoCo, SwAV, SimCLR, SimSiam and SeqCLR which are also contrastive learning-based methods, GMSS achieves better results. 
This is attributed to GMSS having more positive and negative pairs (We set M = 8), and two more pretext tasks, that is, spatial and frequency jigsaw puzzle tasks, which are helpful in learning more discriminant and general EEG emotion representation.
In summary, from the results of Table \ref{tab:3}, in the unsupervised mode, it is observed that GMSS achieves an acceptable results without labels, making it more relevant to practical applications.

To better understand the confusion matrix of GMSS in recognizing different emotions, the unsupervised confusion matrices of all the experiments are displayed in Fig. \ref{Fig.main6}. There are two observations:
\begin{itemize}
	\item[(1)]
	
	For the subject-dependent experiment shown in Fig. \ref{Fig.main6}(1), it is observed that happy is the easiest emotion recognized by SEED dataset. This is also observed in the results of the SEED-IV dataset. For MPED, which contains seven emotions, GMSS shows its superiority when identifying funny, neutral, fear, and anger. In addition, we can find joy is most easily confused with neutral. This may be because joy is more difficult to induce than other emotions.
	\item[(2)]

	From the results of the subject-independent task shown in Fig. \ref{Fig.main6}(2), for SEED, it is obvious that the accuracy of the happy emotion is much higher than neutral and sad, which is similar to the observation in Fig. \ref{Fig.main6}(1). With SEED-IV, we can notice that neutral emotion achieves the highestaccuracy since other emotions such as neutral lead to confusion. For MPED, funny, neutral, and sad emotions are much easier to recognize. It should be noted that, in the cross-subject task, the focus is on the sad emotion, which is difficult to identify from our observation.

\end{itemize}

\begin{figure}
	\centering
	\subfigure[SEED]{\includegraphics[width=0.48\columnwidth]{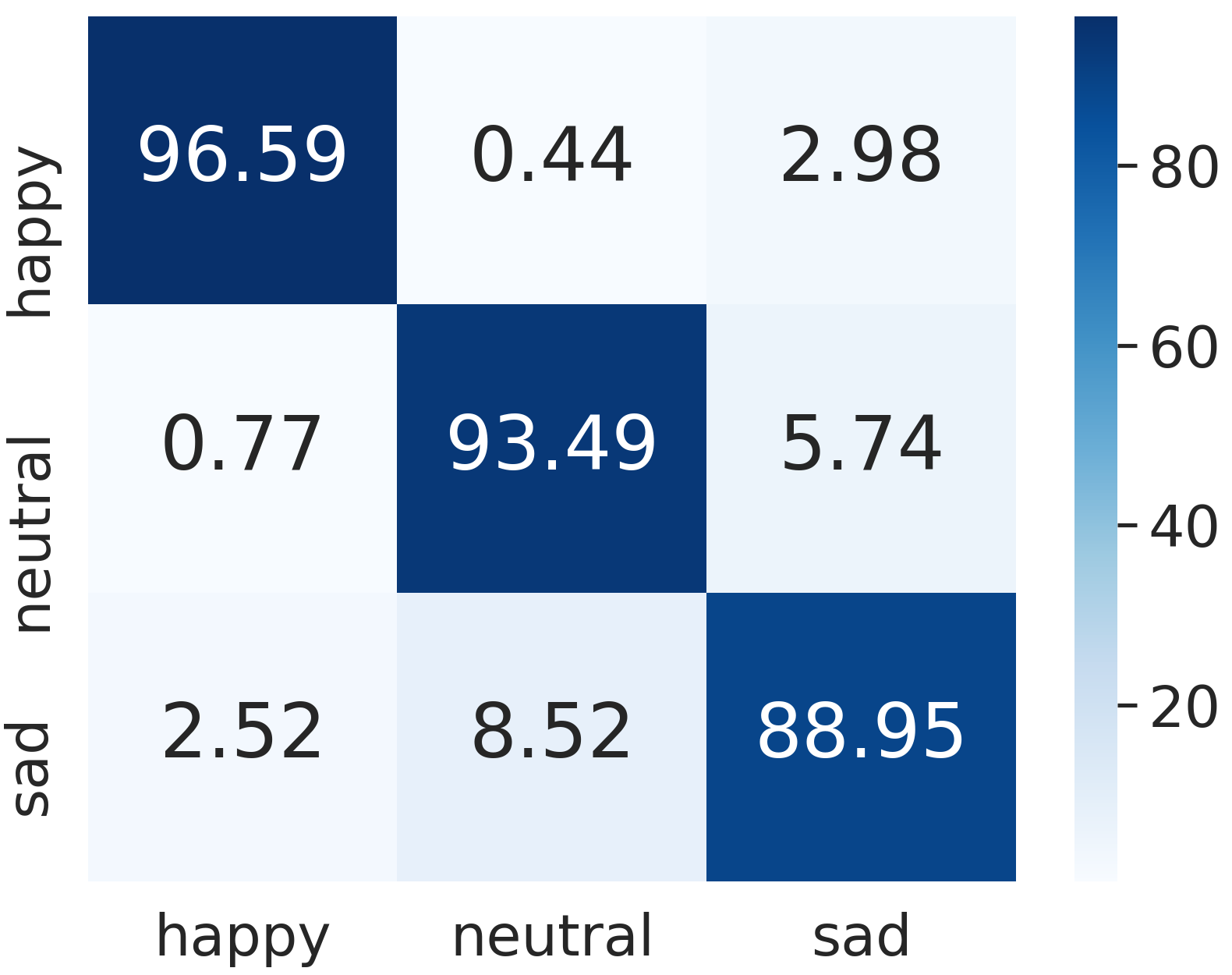}}
	\subfigure[SEED-IV]{\includegraphics[width=0.48\columnwidth]{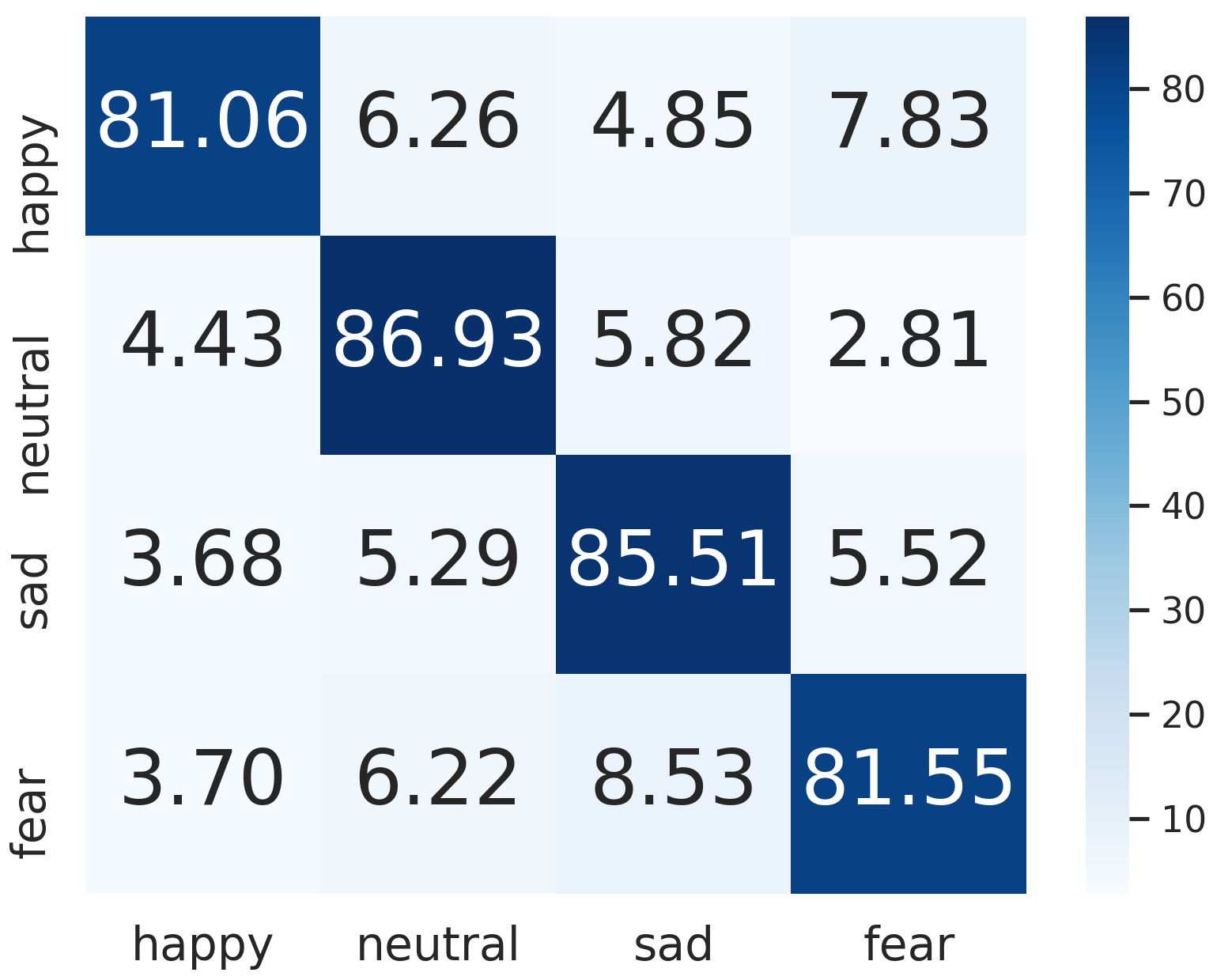}}
	\subfigure[MPED]{\includegraphics[width=0.95\columnwidth]{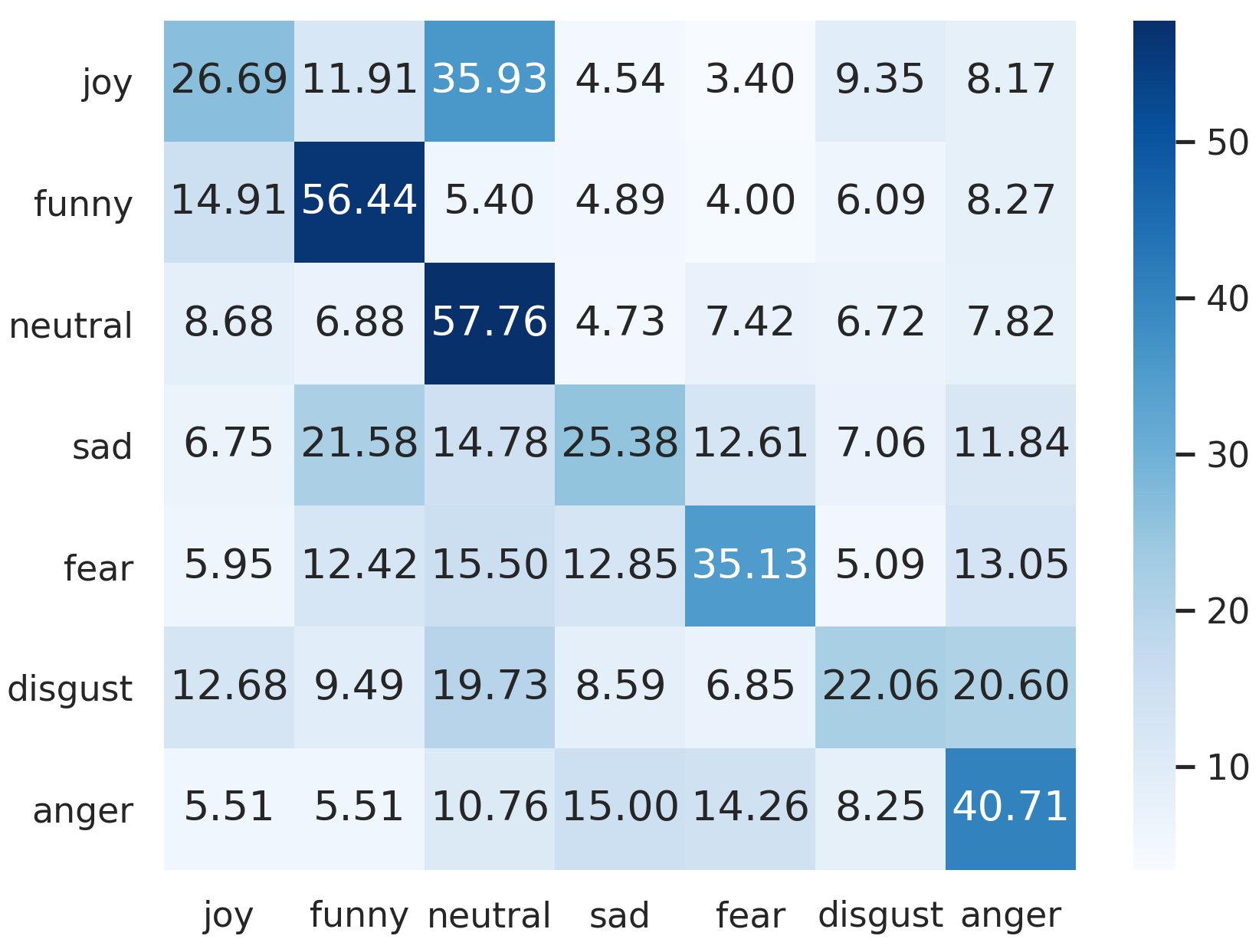}}
	\text{(1) Subject-dependent experimental results }
	\\
	\subfigure[SEED]{\includegraphics[width=0.48\columnwidth]{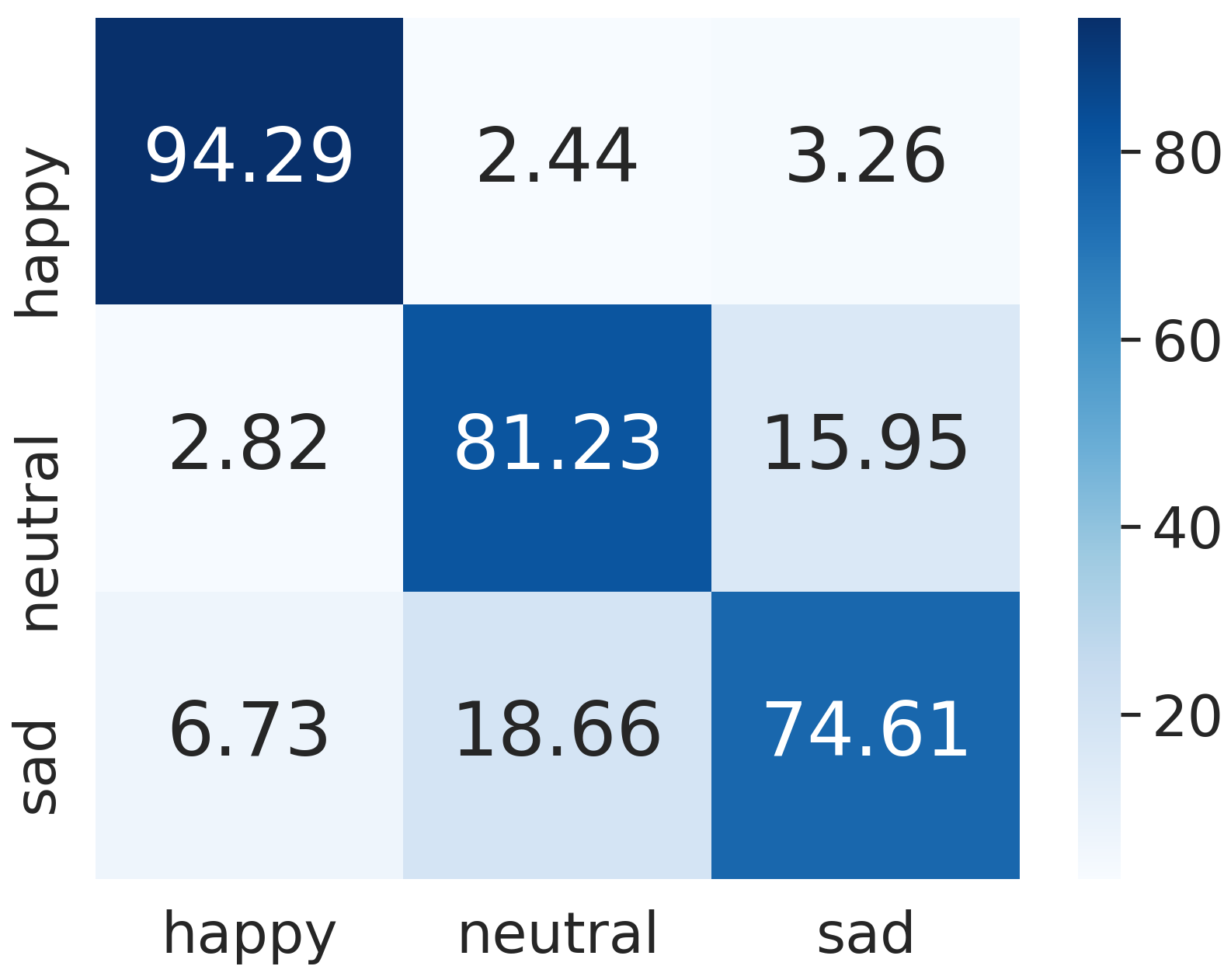}}
	\subfigure[SEED-IV]{\includegraphics[width=0.48\columnwidth]{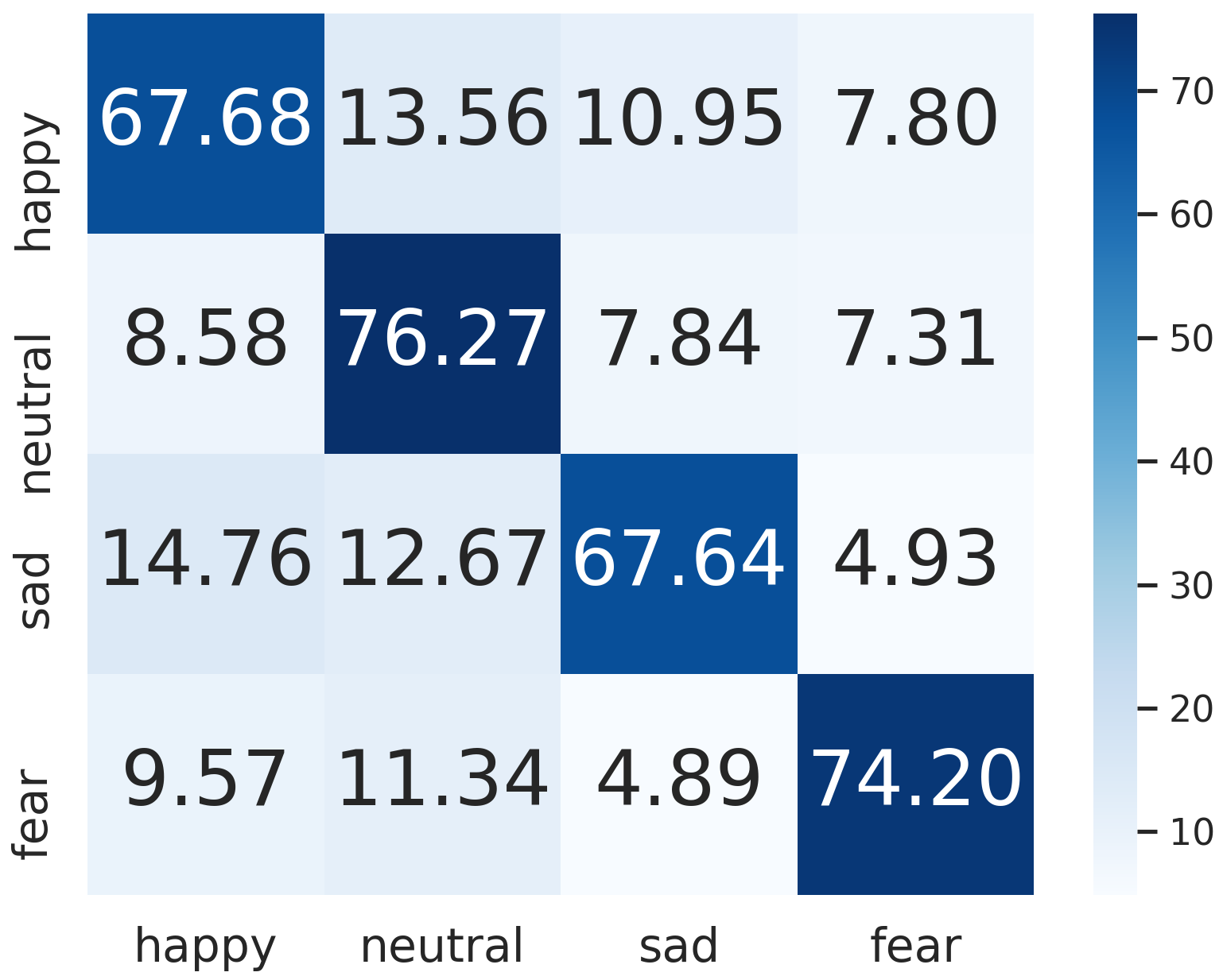}}
	\subfigure[MPED]{\includegraphics[width=0.95\columnwidth]{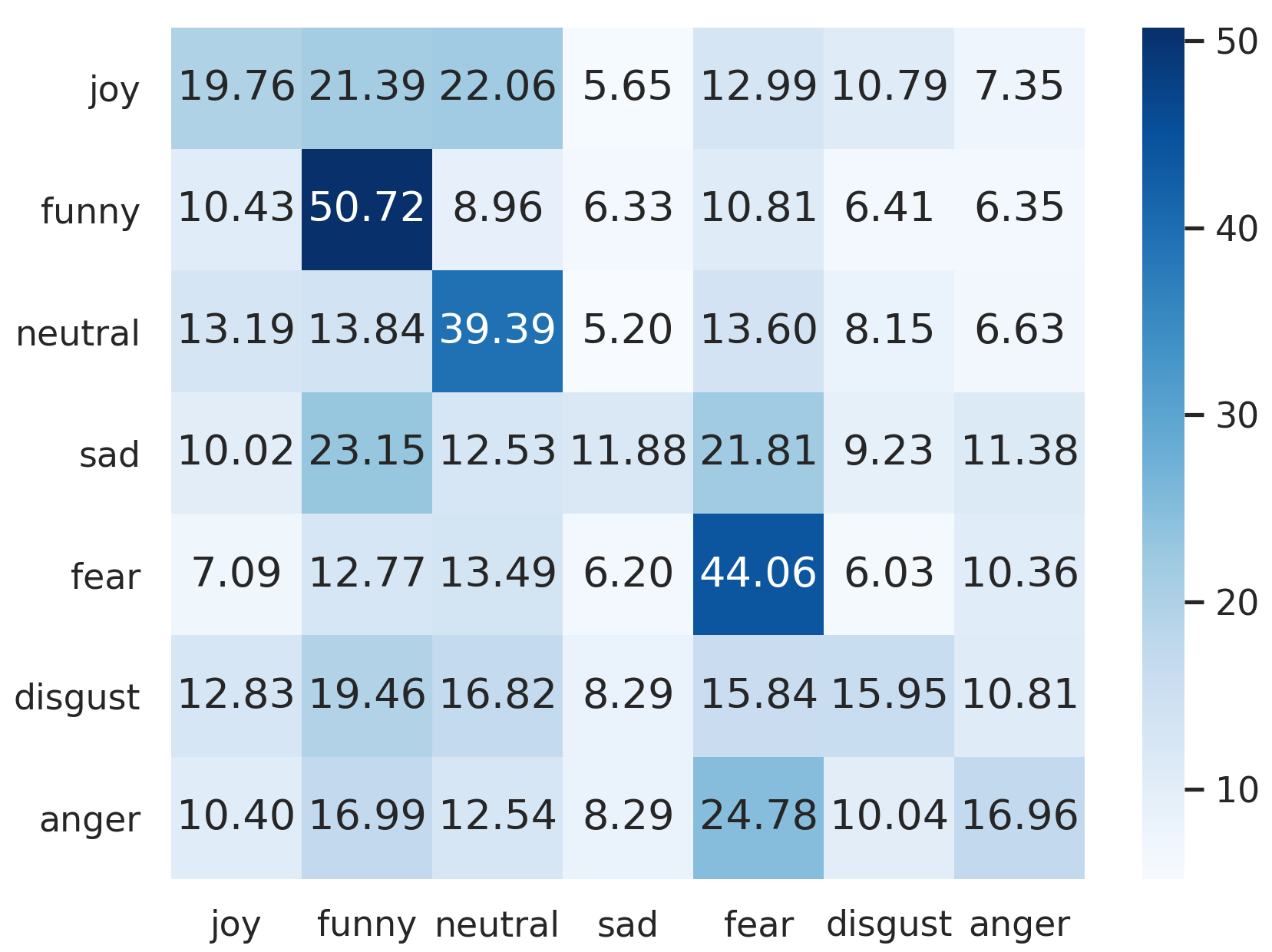}}
	\text{(2) Subject-independent experimental results}
	\\
	
	\caption{ Confusion matrices in supervised mode. (a)-(c) and (d)-(f) are the subject-dependent and subject-independent results on SEED, SEED-IV and MPED datasets, respectively.}
	\label{Fig.main5}
\end{figure}

\subsubsection{Supervised Mode}
In this section, a joint-training strategy is adopted. Based on the self-supervised training approaches, ground-truth emotion labels are used to train the feature extractor simultaneously. To evaluate the advantages of GMSS, the experiments conducted were the same as those of other methods, including linear support vector machine (SVM)\cite{b55}, dynamical graph convolutional neural network (DGCNN)\cite{b11}, regularized graph neural network (RGNN)\cite{b58}, domain adversarial neural networks (DANN)\cite{b21}, bi-hemisphere domain adversarial neural network (BiDANN)\cite{b46}, attention-long short-term memory (A-LSTM)\cite{b41}, and bi-hemispheric discrepancy model for EEG emotion recognition (BiHDM)\cite{b13}. All these methods are representative of previous studies on emotion recognition. Their results are directly quoted or reproduced from the literature to ensure a convincing comparison with the proposed method, and are summarized in Table \ref{tab:2}. 

For the subject-dependent experiments, in Table \ref{tab:2}, it is observed that GMSS attains the best performance on three public EEG emotional datasets compared with all aforementioned methods above. In particular, the results on SEED and SEED-IV, with GMSS are 2.24\% and 7\% higher than those of the most advanced method RGNN. 
Meanwhile, it is also observed that GMSS achieves a performance very close to BiHDM on MPED, that is, 40.16\% vs. 40.34\%. This is because BiHDM is trained not only on the labeled training data but also on the unlabeled testing data. However, the GMSS is trained only on the training data. For a fair comparison, the domain discriminator of BiHDM is ablated and the experiments are conducted on the same input data as GMSS, which is denoted as BiHDM w/o DA. The experimental results show that GMSS improves the classification accuracy by 1.61\% compared with BiHDM w/o DA. Furthermore, GMSS outperforms the BiHDM by 3.36\% and 12.02\% on SEED and SEED-IV datasets, respectively. These results verify that GMSS has a better discrimination capability under subject-dependent experiments. 
Additionally, our GMSS has a considerable running speed. On the SEED dataset of subject-dependent experiments, the average training time and average testing time for one epoch are 3762.7ms and 331.39ms respectively.
Subject-independent experiment are also performed. It is observed that GMSS achieves the SOTA performance on SEED and MPED, which is 1.12\% and 0.22\% higher than the previous best method BiHDM, respectively. Moreover, GMSS achieves a performance close to that of RGNN on SEED-IV, that is, 73.48\% vs. 73.84\%, respectively. However, while RGNN removes its node-wise domain adversarial training component (NodeDAT), that is, training with the labeled training data as well as without the unlabeled testing data, denoted as RGNN w/o DA, the accuracy of RGNN w/o DA is 1.83\% lower than that of GMSS. Furthermore, GMSS outperforms RGNN by 1.22\% on the SEED dataset.
In addition, compared with these advanced methods training without the unlabeled testing data, that is, BiHDM w/o DA and RGNN w/o DA, GMSS is 4.6\%, 1.83\%, and 1.06\% higher, respectively. This indicates that our model can extract more general data representations for different subjects.
Besides, compared with all baselines on all datasets and both experimental protocols, GMSS achieves the lowest standard deviation in accuracy, indicating the excellent discrimination and generalization capability of our model. We argue that the main reason can be attributed to the multi-task framework and self-supervised learning tasks.

Similar to the unsupervised mode, the confusion matrices of all experiments are also applied in the supervised mode to better understand the confusion of GMSS in recognizing different emotions as shown in Fig. \ref{Fig.main5}. There are two observations:
\begin{itemize}
	\item[(1)]
	
	For the results of subject-dependent EEG emotion recognition experiment in Fig. \ref{Fig.main5}(1), the classification accuracy for the three emotions is approximately 90\% for the SEED dataset. In particular, for happy, the accuracy is above 95\%. The happy and neutral emotions are easier to recognize than the sad emotion. For SEED-IV, which contains four emotions, we can notice that the accuracy of all emotions is above 80\%. For MPED, which is a complex dataset that consists of seven types of emotions, it is observed that funny and neutral emotions are much easier to recognize than other emotions. Moreover, for negative emotions, fear and anger are easier to recognize than sad and disgust.

	\item[(2)]
	
	From the results of the subject-independent EEG emotion recognition experiment, for SEED, the happy emotion is much easier to be recognize than neutral and sad emotions. For SEED-IV, neutral and fear emotions are much easier to recognize. For MPED, which is a hard seven classification task, only funny, neutral and fear achieve acceptable results, which suggests that researchers should pay attention to joy, sad, disgust and anger in cross-subject emotion recognition.	
\end{itemize}

\subsection{Discussion}
In this section, the representations of the visualization and ablation studies are presented.

\subsubsection{Representation Visualization}
\begin{figure*}[htb]
	\centering
	\subfigure[Subject-01]{\includegraphics[width=0.4\columnwidth]{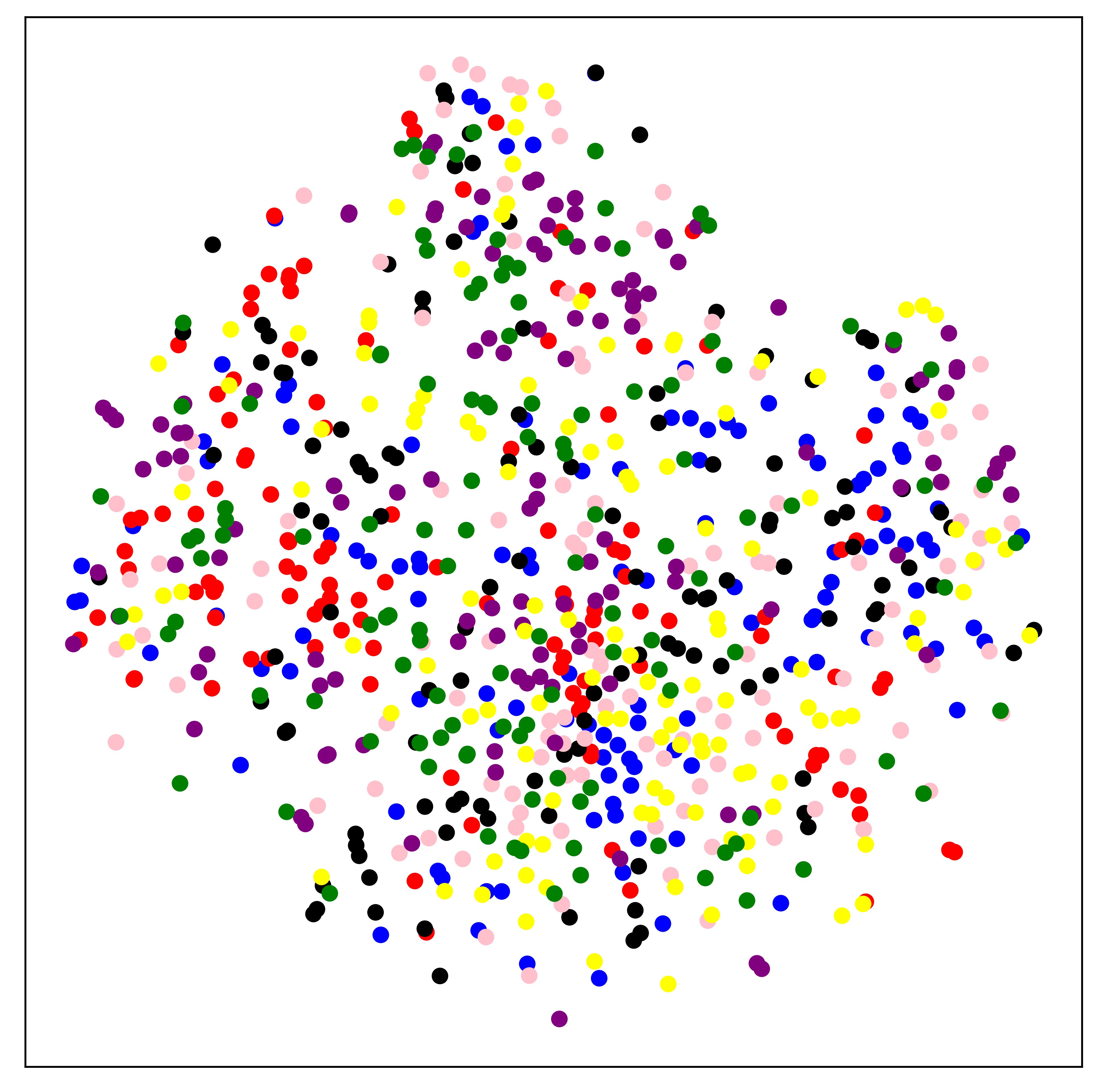}}
	\subfigure[Subject-02]{\includegraphics[width=0.4\columnwidth]{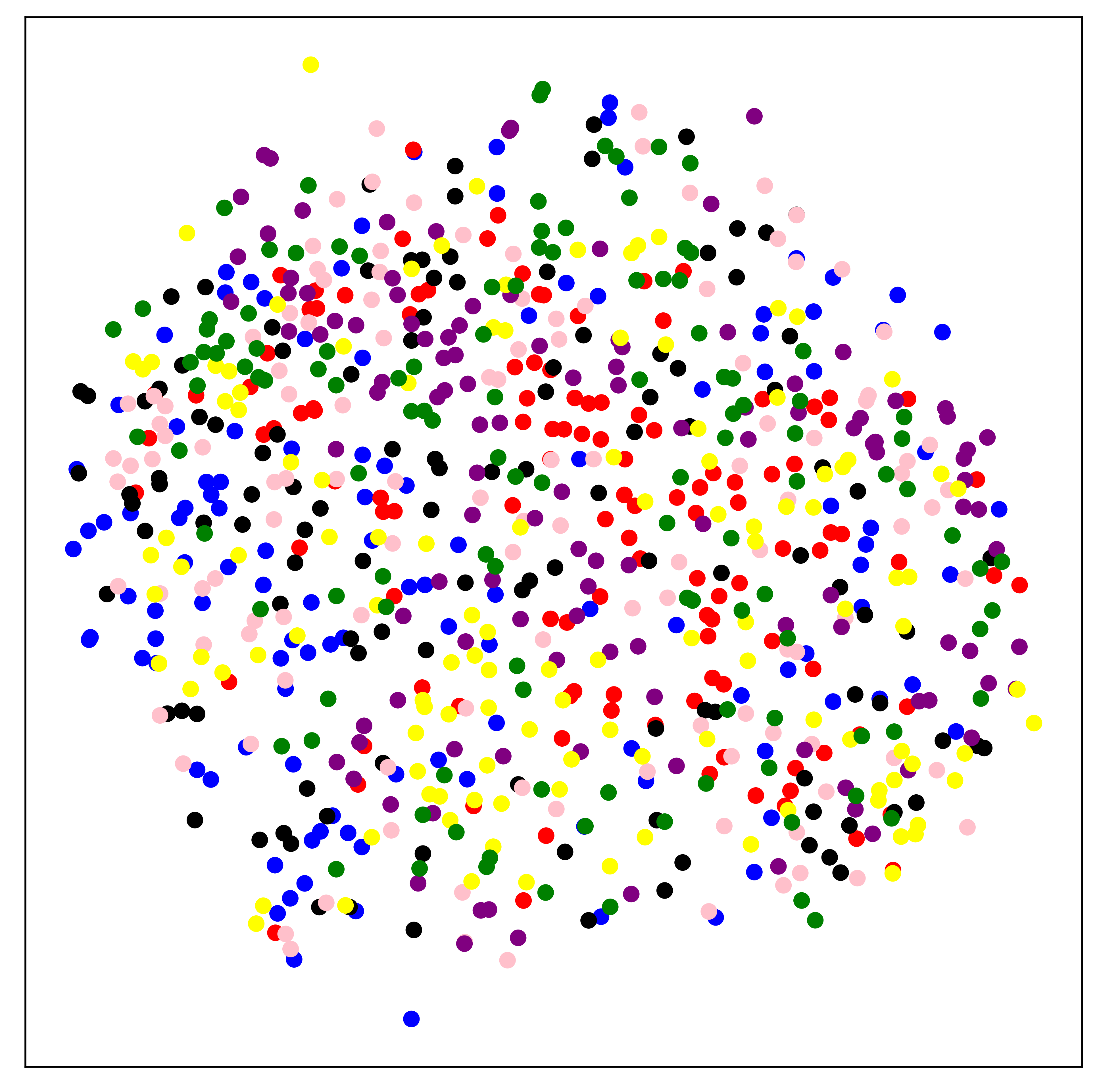}}
	\subfigure[Subject-03]{\includegraphics[width=0.4\columnwidth]{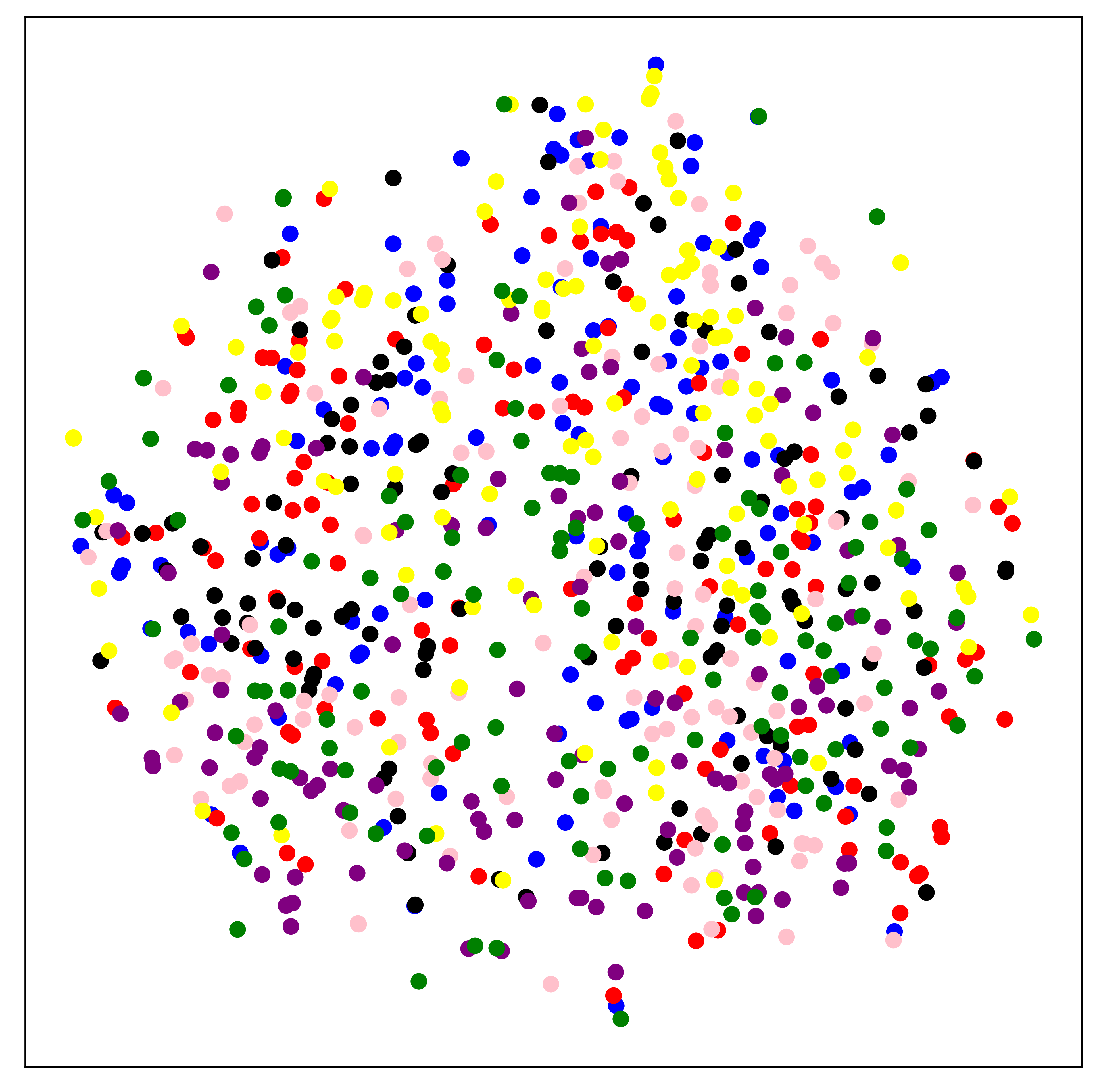}}
	\subfigure[Subject-04]{\includegraphics[width=0.4\columnwidth]{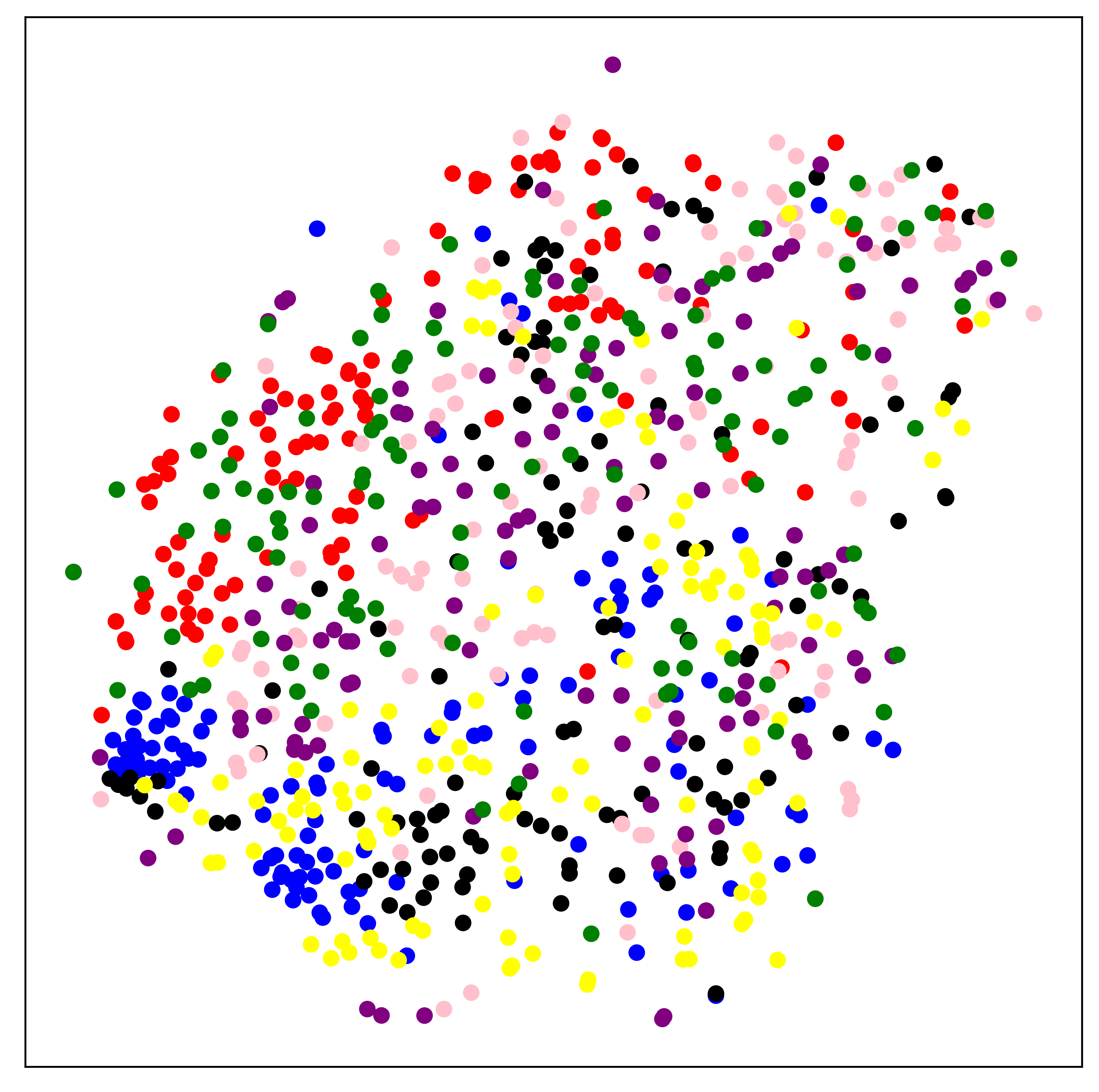}}
	\subfigure[Subject-05]{\includegraphics[width=0.4\columnwidth]{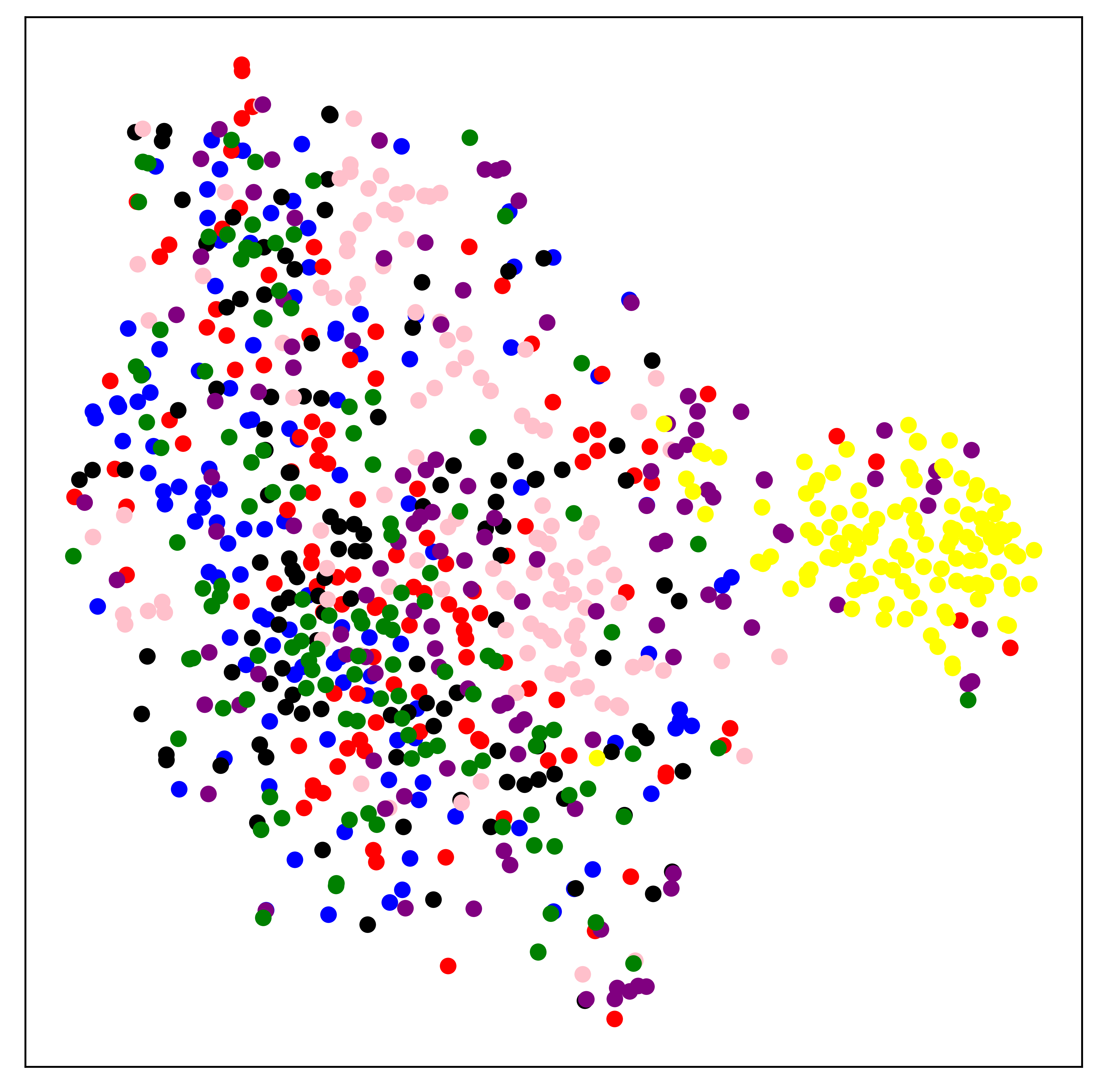}}
	\text{(1) Original EEG emotion data}
	\\
	
	\subfigure[Subject-01]{\includegraphics[width=0.4\columnwidth]{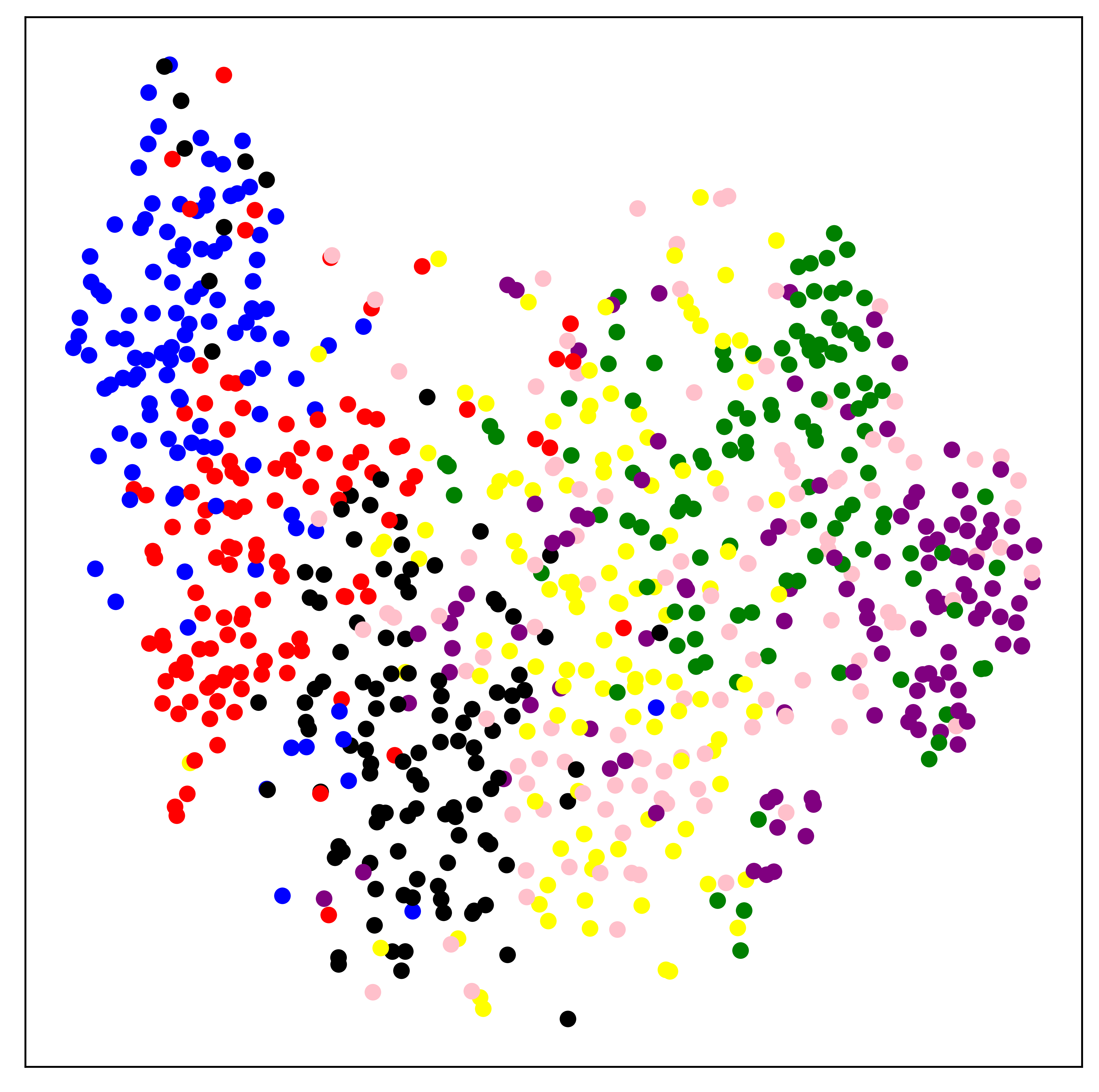}}
	\subfigure[Subject-02]{\includegraphics[width=0.4\columnwidth]{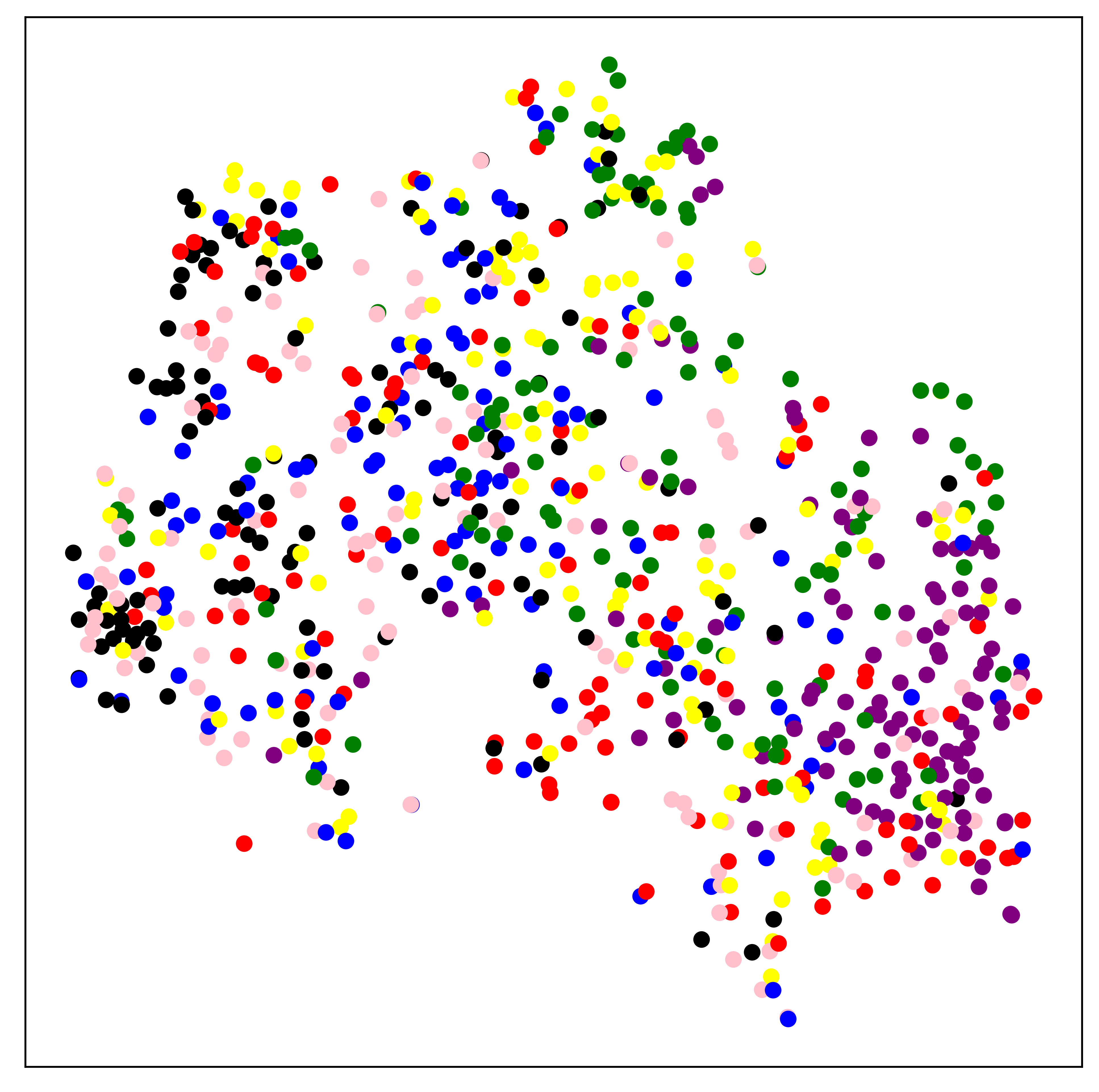}}
	\subfigure[Subject-03]{\includegraphics[width=0.4\columnwidth]{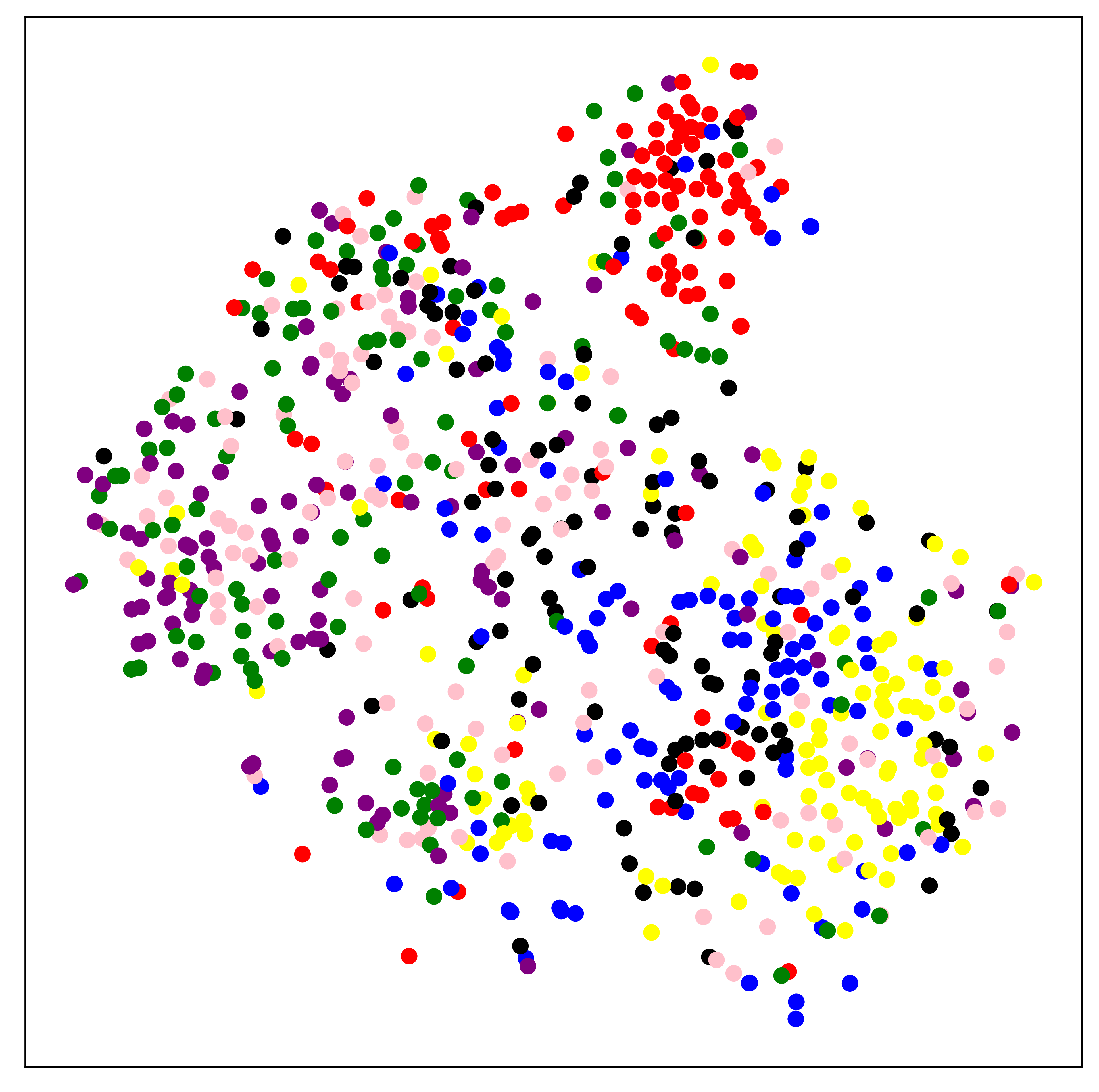}}
	\subfigure[Subject-04]{\includegraphics[width=0.4\columnwidth]{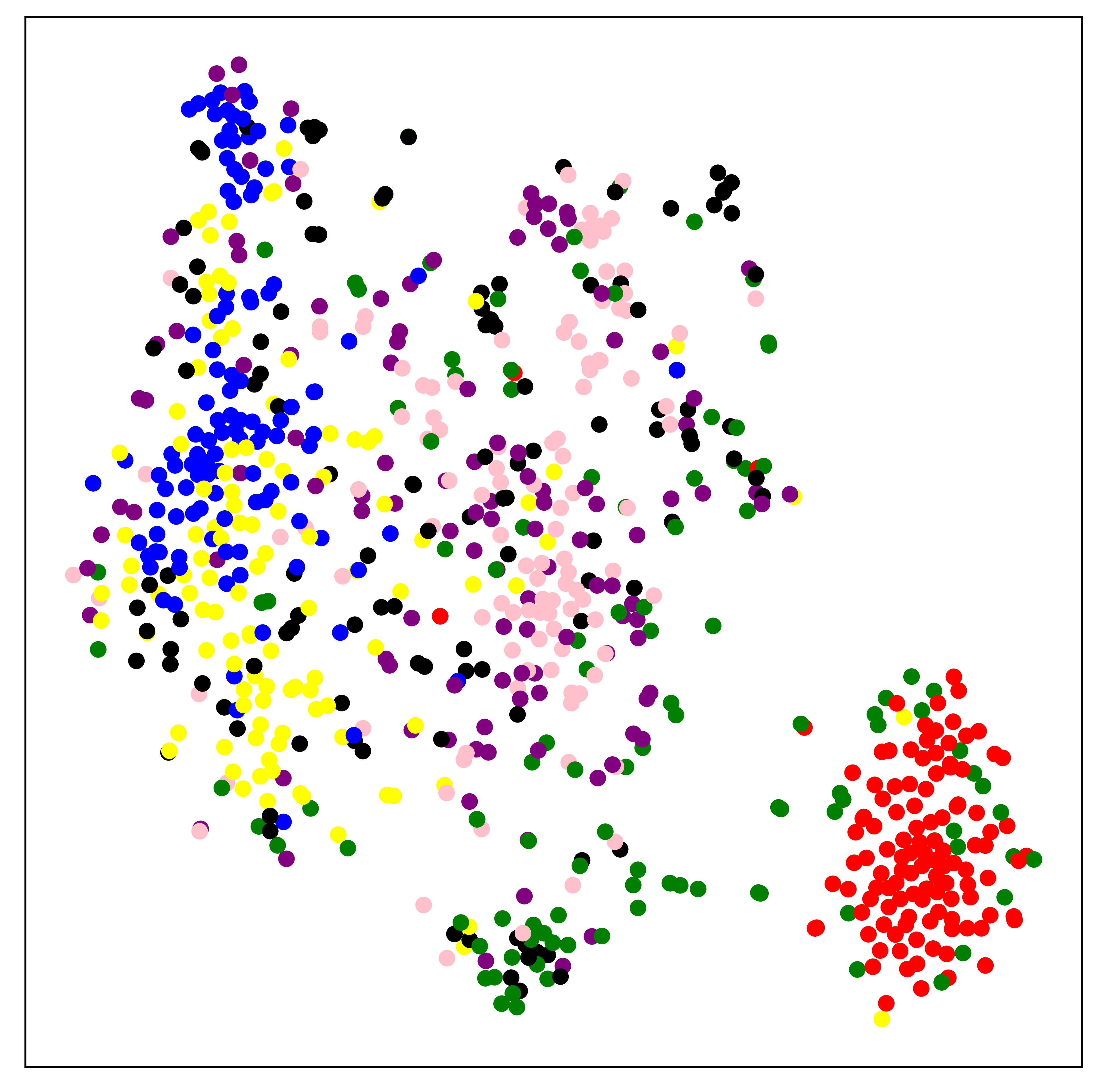}}
	\subfigure[Subject-05]{\includegraphics[width=0.4\columnwidth]{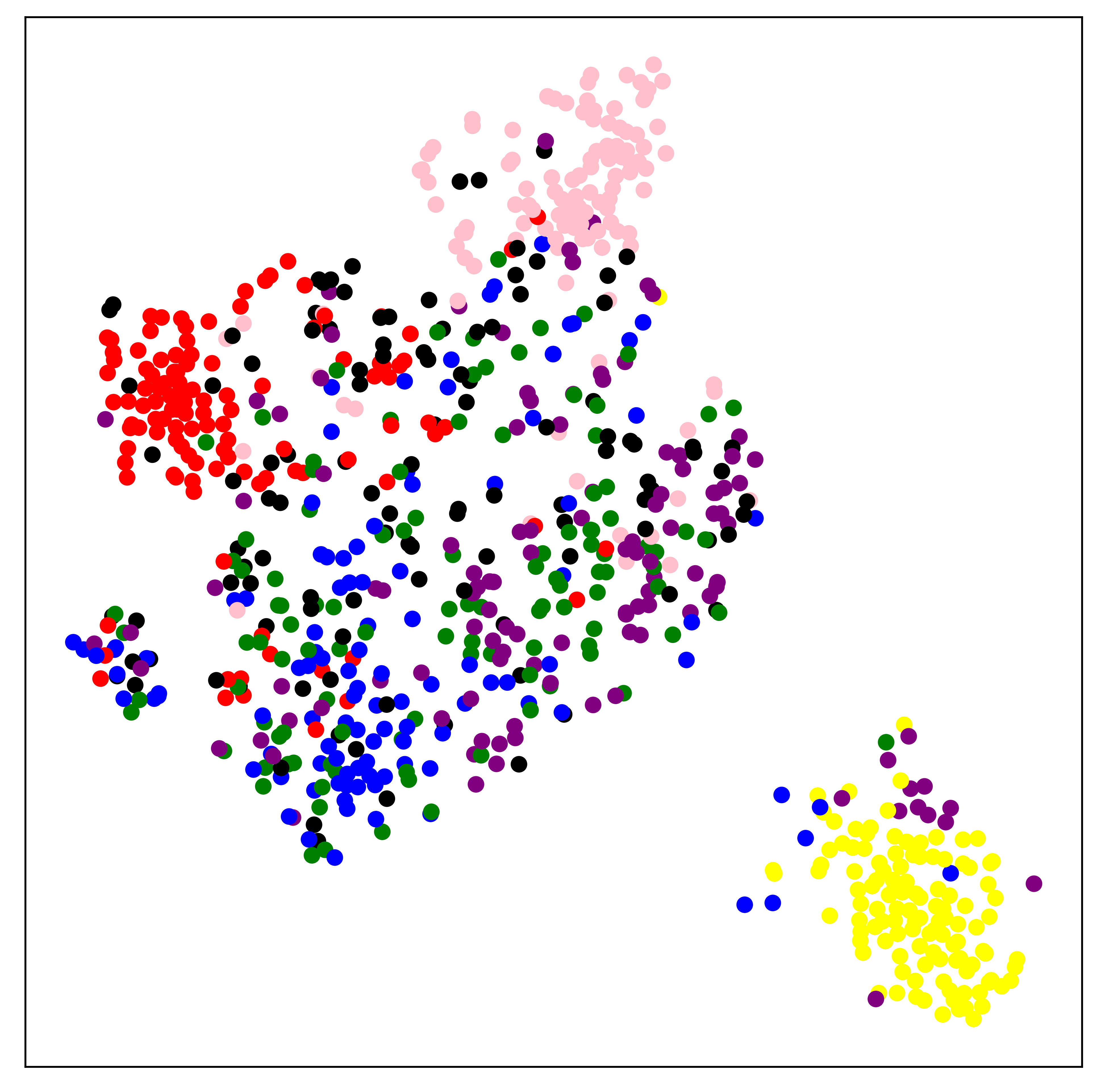}}
	\text{(2) Discriminative representations learned by GMSS}
	\\
	
	\caption{ t-SNE visualization based on original EEG emotion data and discriminative representations learned by GMSS. (a)-(e) are the distributions of original EEG emotion data before being fed into network; (f)-(j) are the learned discriminative representations by GMSS. Blue, red, yellow, green, purple, black and pink dots denote joy, funny, neutral, sad, fear, disgust and anger emotions, respectively.}
	\label{Fig.main9}
\end{figure*}

To verify the discriminating ability of GMSS, the features obtained by GMSS on the MPED dataset are visualized.
Fig. \ref{Fig.main9} shows the representation visualization of the subject-dependent experiment in supervised mode using t-distributed stochastic neighbor embedding (t-SNE)\cite{b59} on the MPED dataset. 
As shown in Fig. \ref{Fig.main9}(1),  it is difficult to separate the different classes from the original EEG data. However, for the learned EEG representation in Fig. \ref{Fig.main9}(2), for the same emotion clusters, there are clear borders between different emotions, which verify that GMSS can discriminate features for EEG emotion recognition. Moreover, it is observed that the funny is more distinguishable than other emotions. This may be because funny induced more easily.
In addition, comparing with Fig. \ref{Fig.main9}(1) and Fig. \ref{Fig.main9}(2), it is observed that GMSS has the potential to clarify the borders of various emotions and brings the same emotions closer together in feature space.

\subsubsection{Ablation study}
To assess the contribution of each essential pretext task in our model, experiments are conducted with the ablated GMSS models in both unsupervised and supervised modes.
The ablation research verifies the influence of each pretext task and the combination of multiple tasks on the performance of EEG emotion recognition.
In Table \ref{tab:4}, the results are presented for the subject-dependent experiments in both unsupervised and supervised modes. In the unsupervised mode, GMSS-S, -F, and -C denote that the only spatial jigsaw puzzle task, frequency jigsaw puzzle task, and contrastive learning task are taken into consideration in the ablation model. Furthermore, GMSS-SF, -SC, -FC denote the spatial and frequency jigsaw puzzle tasks, spatial jigsaw puzzle and contrastive learning tasks, frequency jigsaw puzzle and contrastive learning tasks respectively taken into consideration by the ablation model, simultaneously. Similarly, in the supervised mode, GMSS-F, -C, -SF, -SC, and -FC represent the same ablation methods but are trained on the ground-truth emotion labels instead.

In the case of one self-supervised pretext task, GMSS-S achieves the best performance on four out of six results.
This indicates that the spatial jigsaw puzzle task is extremely helpful in improving the discrimination of EEG emotional signals. 
Moreover, GMSS-F achieves the best performance on two out of six results, which implies that the frequency jigsaw puzzle task is helpful as well. The above results demonstrate that only one jigsaw puzzle task could improve the ability to distinguish EEG emotion signals.
In the case of two tasks, GMSS-SF achieves the best performance on all datasets except MPED in the supervised mode, which is slightly lower than that of GMSS-SC. This further proves the effectiveness of the jigsaw puzzle task. In addition, compared with the corresponding results of only one task, the combination of the two tasks improve the accuracy of emotion recognition. This indicates that the three self-supervised tasks that were proposed are relevant and can promote model learning and more discriminative emotional representation.  
Furthermore, we can see that GMSS adopts all pretext tasks, achieving the best performance. This proves the effectiveness of our graph-based multi-task self-supervised learning framework.

\begin{table*}[htb]
	\centering
	\caption{Ablation study of subject-dependent classification accuracy (mean/std)  for unsupervised mode and supervised mode on SEED, SEED-IV, and MPED datasets}\label{tab:4}
	\begin{threeparttable}
		\begin{tabular}{ccccccc}
			\toprule
			\multirow{2}{*}{Ablation Models} & \multicolumn{2}{c}{SEED}        & \multicolumn{2}{c}{SEED-IV}      & \multicolumn{2}{c}{MPED}        \\
			& unsupervised & supervised & unsupervised & supervised & unsupervised & supervised \\
			\midrule
			GMSS-S & \textbf{86.43/09.36}      & \textbf{88.82/08.81}         & \textbf{63.29/16.50}     & 79.01/16.13          & 31.91/06.09     & \textbf{35.82/06.17}         \\
			
			GMSS-F & 84.84/10.68      & 86.75/08.64          & 62.31/16.24     & \textbf{79.56/14.31}         & \textbf{33.32/06.45}    & 34.98/06.13          \\
			
			GMSS-C & 84.14/10.65       & 85.92/09.78         & 59.77/17.64     & 77.68/15.02          & 32.28/06.07     & 35.59/05.99          \\
			
			\midrule
			
			GMSS-SF & \textbf{88.24/09.77}      & \textbf{94.98/09.34}          & \textbf{64.21/14.92}     & \textbf{84.54/14.30}          & \textbf{33.81/05.67}     & 37.78/05.95          \\
			
			GMSS-SC & 86.81/10.37      & 93.94/09.57          & 62.66/17.47     & 83.42/11.83        & 32.42/06.42     & \textbf{38.06/05.65}         \\
			
			GMSS-FC & 86.35/10.15      & 92.93/08.29         & 62.83/17.29     & 83.83/12.49          & 33.65/06.66     & 37.11/05.97          \\
			
			\midrule
			
			GMSS & \textbf{89.18/09.74}       & \textbf{96.48/04.63}          & \textbf{65.61/17.33}     & \textbf{86.37/11.45}          & \textbf{34.81/06.88}     & \textbf{40.16/06.08}         \\
			\bottomrule
		\end{tabular}

	\end{threeparttable}
\end{table*}

\begin{figure}[htb] 
	
	\centering 
	\includegraphics[width=.48\textwidth]{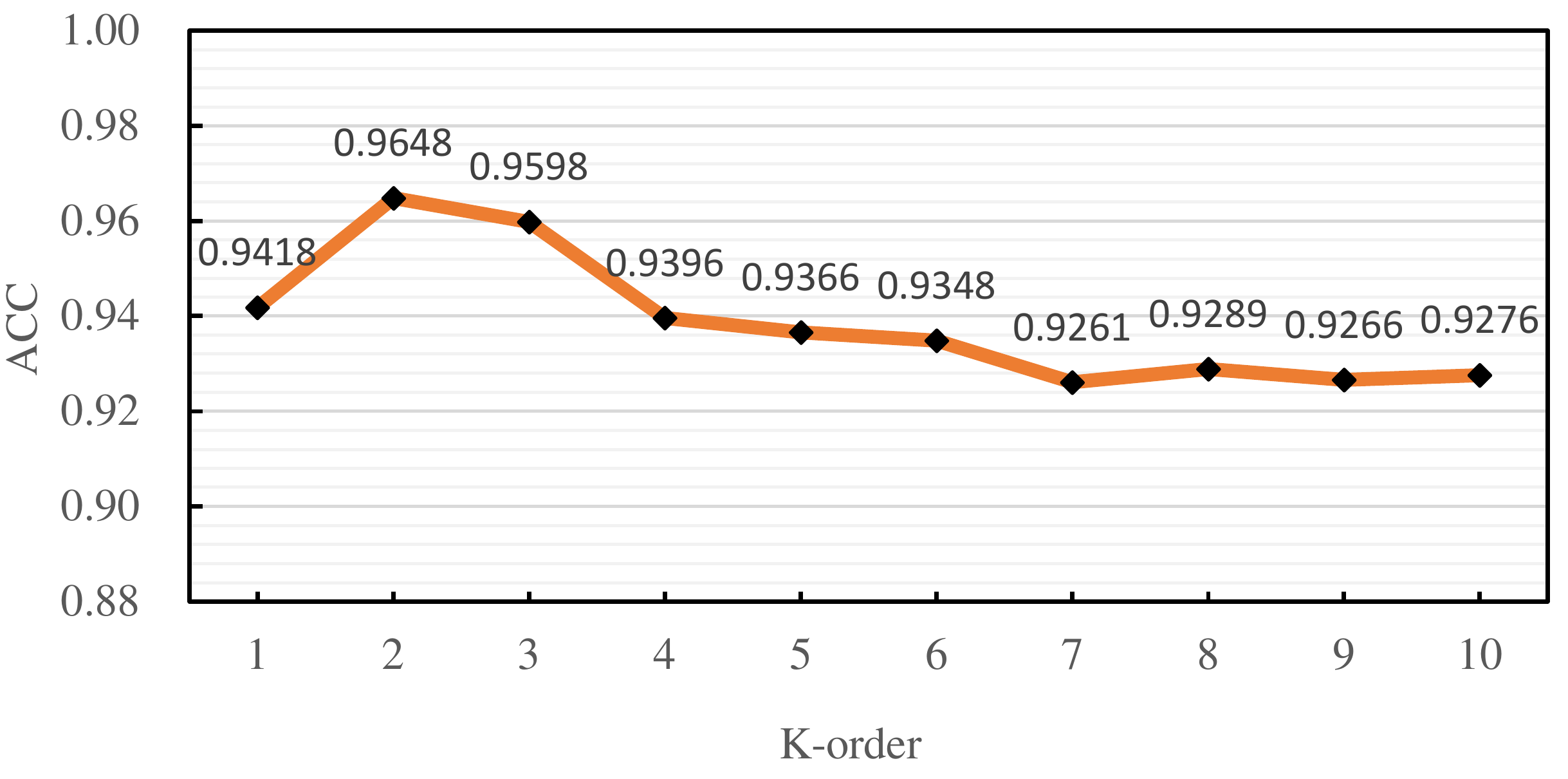}
	\caption{Experiment results based on different Chebyshv filter sizes.} 
	\label{Fig.chart} 
\end{figure}

\subsubsection{Parameter analysis - Chebyshev filter size}

As a hyper-parameter, Chebyshv filter size $K$, namely, $K$-order neighbor, will impact the performance of  EEG emotion recognition. Thus, in this section, we conduct additional experiment to analyze the results of different Chebyshv filter size $K$ on SEED dataset. Here we set $K$ = 1, 2, ..., 10 separately. And the results are shown in Fig. \ref{Fig.chart}. It is obvious that GMSS achieves the best performance when $K=2$ . When $K$ is greater than 2, the performance of the model has a relatively noticeable downward trend. When $K$ is greater than 4, it tends to be stable gradually. We attribute the decline to the influence of over-smoothing. 
\color{black}

\section{Conclusion}
In this paper, a graph-based multi-task self-supervised learning model is proposed for EEG emotion recognition. Our model is inspired by the multi-task learning theory and self-supervised learning theory, which combines different self-supervised tasks to improve model generalization and the ability to recognize EEG emotional signals. Several self-supervised tasks assist in improving the resilience of the model to emotion noise labels. The spatial pattern of EEG emotion signals is studied through the spatial jigsaw puzzle task. To reveal the intrinsic frequency bands for EEG emotion recognition, the frequency jigsaw puzzle task is employed, and the feature space is further standardized by the contrastive learning tasks.
The experimental results validate the effectiveness of the proposed model. In future work, multi-task self-supervised learning will be further investigated to explore how to further improve EEG emotion recognition.

\bibliographystyle{IEEEtran}
\bibliography{reference}

\end{document}